\documentclass[sn-mathphys,Numbered]{sn-jnl}

\usepackage{graphicx}%
\usepackage{multirow}%
\usepackage{amsmath,amssymb,amsfonts}%
\usepackage{amsthm}%
\usepackage{mathrsfs}%
\usepackage[title]{appendix}%
\usepackage{color, xcolor}%
\usepackage{textcomp}%
\usepackage{manyfoot}%
\usepackage{booktabs}%
\usepackage{algorithm}%
\usepackage{algorithmicx}%
\usepackage{algpseudocode}%
\usepackage{listings}%
\usepackage{makecell}
\usepackage[section]{placeins}
\usepackage{siunitx}
\usepackage{lineno}
\usepackage[utf8]{inputenc}
\usepackage[T1]{fontenc}
\usepackage{geometry}

\usepackage{listings}%
\usepackage[section]{placeins}
\usepackage{lineno}
\usepackage{hyperref}
\DeclareMathOperator{\diag}{diag}

\usepackage{geometry}

\usepackage{soul}
\soulregister{\cite}7 
\soulregister{\citep}7 
\soulregister{\citet}7 
\soulregister{\ref}7 
\soulregister{\pageref}7 

\theoremstyle{thmstyleone}%
\theoremstyle{thmstyletwo}%

\theoremstyle{thmstylethree}%

\makeatletter
\newcommand*{\rom}[1]{\expandafter\@slowromancap\romannumeral #1@}
\makeatother

\begin{document}

\title[Article Title]{Physics-informed generative real-time lens-free imaging}


\author*[1,2,3]{\fnm{Ronald B.} \sur{Liu}}\email{ronald.liu@ed.ac.uk}

\author[1]{\fnm{Zhe} \sur{Liu}}\email{zz.liu@ed.ac.uk}

\author[3]{\fnm{Max G.A.} \sur{Wolf}}\email{max.wolf@kuleuven.be}

\author[2]{\fnm{Krishna P.} \sur{Purohit}}\email{s1785889@sms.ed.ac.uk}

\author[3]{\fnm{Gregor} \sur{Fritz}}\email{gregor.fritz@kuleuven.be}


\author[2]{\fnm{Yi} \sur{Feng}}\email{yi.feng@ed.ac.uk}

\author[2]{\fnm{Carsten G.} \sur{Hansen}}\email{Carsten.G.Hansen@ed.ac.uk}

\author*[2]{\fnm{Pierre O.} \sur{Bagnaninchi}}\email{pierre.bagnaninchi@ed.ac.uk}

\author*[3]{\fnm{Xavier} \sur{Casadevall i Solvas}}\email{xevi.casadevall@kuleuven.be}

\author*[1]{\fnm{Yunjie} \sur{Yang}}\email{y.yang@ed.ac.uk}

\affil[1]{\orgdiv{SMART Group, Institute for Imaging, Data and Communications, School of Engineering}, \orgname{University of Edinburgh}, \orgaddress{\country{UK}}}

\affil[2]{\orgdiv{Institute Regeneration and Repair, College of Medicine and Veterinary Medicine}, \orgname{University of Edinburgh}, \orgaddress{\country{UK}}}

\affil[3]{\orgdiv{Biomimetics Group, Department of Biosystems}, \orgname{Catholic University of Leuven}, \orgaddress{\country{Belgium}}}


\abstract{Advancements in high-throughput biomedical applications require real-time, large field-of-view (FOV) imaging. While current 2D lens-free imaging (LFI) systems improve FOV, they are often hindered by time-consuming multi-position measurements, extensive data pre-processing, and strict optical parameterization, limiting their application to static, thin samples. To overcome these limitations, we introduce GenLFI, combining a generative unsupervised physics-informed neural network (PINN) with a large FOV LFI setup for straightforward holographic image reconstruction, without multi-measurement. GenLFI enables real-time 2D imaging for 3D samples, such as droplet-based microfluidics and 3D cell models, in dynamic complex optical fields. Unlike previous methods, our approach decouples the reconstruction algorithm from optical setup parameters, enabling a large FOV limited only by hardware. We demonstrate a real-time FOV exceeding 550 mm$^2$, over 20 times larger than current real-time LFI systems. This framework unlocks the potential of LFI systems, providing a robust tool for advancing automated high-throughput biomedical applications.}

\maketitle

\section*{Introduction} \label{sec1}
\textit{High-throughput} is a critical feature of automated biomedical imaging systems, reflecting the efficiency of analysis in terms of field-of-view (FOV), resolution, and imaging speed. Most current biomedical imaging modalities are based on light microscopy, in which light is transmitted through or reflected from the sample via lenses. Significant advancements have been made in lens-based microscopy, including improvements in contrast \cite{zernikePhaseContrastNew1942}, specificity \cite{menziesSensitivitySpecificityAnalysis1996,SensitivitySpecificityEpiluminescence}, optical sectioning \cite{agardOpticalSectioningMicroscopy1984,conchelloOpticalSectioningMicroscopy2005}, and super-resolution techniques \cite{nehmeDeepSTORMSuperresolutionSinglemolecule2018,choongQuantumAnnealingSingle2023,wangGlobalAlignedStructured2023,ouyangDeepLearningMassively2018,qiaoEvaluationDevelopmentDeep2021}. However, the FOV in these systems remains constrained by magnification limits and lens aberrations.
To image larger areas, current high-throughput platforms—such as microscopes, slide scanners, and plate readers—typically rely on mechanical \textit{scanning}, which involves repeated adjustments of the lens position \cite{zhangLongtermMesoscaleImaging2024}. Recent developments, such as phase-mask-based diffuse cameras \cite{xueSingleshot3DWidefield2020, adamsVivoLenslessMicroscopy2022, antipaDiffuserCamLenslessSingleexposure2018, tianGEOMScopeLargeFieldofView2021}, aim to overcome the limitations of conventional optics. While promising, these systems often rely on fluorescent biomarkers and face challenges in real-time imaging due to computational delays in image reconstruction. Other approaches, such as integrating multiple lenses \cite{thomsonGigapixelImagingNovel2022, harfoucheImagingMultipleSpatial2023a, huMetalensArrayMiniaturized2024}, face challenges in achieving an optimal balance between FOV and resolution, synchronization for dynamic imaging and optical crosstalks.
These inherent limitations hinder real-time, label-free imaging of large biological specimens, such as extensive tissue sections (e.g., tracking metastatic cancer across microvessels), entire model organisms (e.g., zebrafish), and complex 3D models (e.g., spheroids, organoids, and organs-on-chips).

To overcome these limitations, a more straightforward solution, \textit{lens-free imaging (LFI)} systems have been developed. LFI, a form of computational imaging, captures diffraction and projection patterns (\textit{i.e., }holograms) directly with the image sensor and reconstructs focused images using computational algorithms, bypassing the need for conventional lenses. The largest achievable FOV of in-line LFI systems is only limited by the image sensor's size \cite{greenbaumImagingLensesAchievements2012}. Although deep learning (DL) methods \cite{huangSelfsupervisedLearningHologram2023, chenEFINEnhancedFourier2023, chenDeepLearningbasedHologram2023} have provided holographic reconstruction (re-focusing) qualities comparable to lens-based microscopes, enabling generalization to diverse sample types, current 2D LFIs still face several intrinsic limitations in actual imaging tasks. (1) Achieving high-resolution resolvable holograms often requires multi-time measurements, like horizontal shifting for sub-pixel super-resolution \cite{greenbaumImagingLensesAchievements2012, zhang3DImagingOptically2017}, or multi-wavelength illumination \cite{marienColorLensfreeImaging2020} (illustrated in supplementary Fig.S2). Also, although current LFI systems \cite{wangEHoloNetLearningbasedEndtoend2018, jayapalaLensfreeHolographicImaging, zhangResolutionAnalysisLensFree2020, barkleyHolographicMicroscopyPython2018, castanedaPyDHMPythonLibrary2022, changPicture3DHolography2023} offer a degree of flexibility in the depth of field through back-propagation, they often require z-stack holograms from multi-position measurements, such as multi-height phase recovery \cite{greenbaumWidefieldComputationalImaging2014, rivensonPhaseRecoveryHolographic2018a, wuBrightfieldHolographyCrossmodality2019}. This multi-measurement requirement hinders the potential for single-shot capture in real-time imaging and limits the applicability of current LFI systems to static samples. 
(2) Current reconstruction algorithms hinge upon resolvable holograms based on free space propagation modeling by angular spectrum method \cite{goodman2005fourier}. In real-world scenarios, this necessitates a clean, non-overlapping, parametrizable wave diffraction optical field, along with prior knowledge of the optical setup and electromagnetic wave characteristics  \cite{greenbaumImagingLensesAchievements2012, greenbaumWidefieldComputationalImaging2014, huangSelfsupervisedLearningHologram2023, ozcanSingleShotAutofocusingMicroscopy2023}. It requires precise control over the sample to a thin layer of material that modulates the incident plane wave with a recoverable phase, allowing it to be parameterized and resolved. This angular spectrum method is significantly challenging when encountering dense or 3D volumetric samples that produce a dynamic complex optical field, as depicted in Fig.\ref{fig-overview}\textbf{b}. (3) To acquire high-resolution, resolvable holograms, current LFI setups \cite{greenbaumWidefieldComputationalImaging2014, alma9993244227901488} mostly use sensors with small pixel sizes. However, due to light intensity and manufacturing limitations, these sensors have small sensor sizes (\(< 20\, mm^2 \)) \cite{alma9993244227901488}, which, in return, further hampers FOV advancements in LFI.

In terms of algorithm, most of the current DL methods in computational imaging can be summarized into three categories: (1) Supervised learning models; these models are based on super-resolution (SR) denoising models, \textit{e.g.,} RCAN-based approaches \cite{zhangImageSuperResolutionUsing2018, weigertContentawareImageRestoration2018, chenThreedimensionalResidualChannel2021}, AdaSR \cite{zhangAdaptiveSuperresolutionEnabled2021}, and physics-based virtual-scanning light-field microscopy (VsLFM) \cite{luVirtualscanningLightfieldMicroscopy2023}. Although these methods can establish a sound incorporation of explicit physical models, \textit{e.g.,} phase imaging \cite{kandelPhaseImagingComputational2020}, these supervised learning approaches rely on large-scale paired ground truth, which is labor-intensive and experimentally impractical. (2) Self-supervised learning, which leverages large-scale pre-training followed by fine-tuning or direct deployment for downstream reconstruction tasks, offers data efficiency. However, when applied to LFI \cite{huangSelfsupervisedLearningHologram2023}, both (1) supervised and (2) self-supervised models demand paired datasets or datasets through a precise optical parameterization to ensure consistency between pre-training and reconstruction, as shown in Fig.\ref{fig-overview}\textbf{c}. This demands stringent control over optical setups and static samples, as we previously mentioned. (3) Unsupervised learning, typically based on the generative image-to-image translation models (I2I), \textit{e.g.,} CycleGAN  \cite{zhuUnpairedImagetoImageTranslation2017} and CUT  \cite{park2020contrastive, shenNoninvasiveNonlinearOptical2024}, can process unpaired datasets based on images' semantic relationships between two image domain distributions. Such semantic features are difficult to extract when encountering fuzzy objects, \textit{e.g.,} holograms or fluorescence.  In LFI, the current CycleGAN-based algorithm can only process distinguishable objects \cite{scherrerHolographicReconstructionEnhancement2022,caoLabelfreeIntraoperativeHistology2023}, and lack of physics prior knowledge. This often causes the style-based image transformation lose its physical consistency and reliability. Furthermore, semantic I2I models cannot extract the latent features of unaligned datasets \cite{xieUnalignedImagetoImageTranslation2021,zhuUnpairedImagetoImageTranslation2017}. This necessitates aligned ROIs across different domains, \textit{i.e.,} hologram and microscopy \cite{scherrerHolographicReconstructionEnhancement2022}, which is not feasible for dynamic objects.

Here, we introduce a straightforward unsupervised framework, GenLFI, to overcome these limitations. GenLFI simplifies the imaging process by eliminating the need for data pre-processing, annotation, and optical field parameterization, directly reconstructing holograms using a physics-informed neural network (PINN), namely LensGAN. LensGAN autonomously learns image features from unpaired and unaligned holograms and bright-field microscopy, generating high-resolution reconstructions. This process is guided by a learning bias incorporating physical prior knowledge of diffraction and projection. Through a series of imaging experiments, we demonstrate the advancement of the GenLFI over existing 2D LFI systems in four key aspects: (1) To enable \textit{real-time} imaging, GenLFI processes \textit{single-shot} \textit{large FOV} holograms, eliminating multi-position measurements and surpassing current hologram reconstruction algorithms in interference speed \cite{huangSelfsupervisedLearningHologram2023,rivensonPhaseRecoveryHolographic2018a}.  (2) The reconstruction algorithm, LensGAN, is an unsupervised learning network that processes unpaired and unaligned training datasets, facilitating feasible large-scale data acquisition. (3) Eliminating the optical field modeling by existing LFI methods, GenLFI is capable of imaging various complex, dense \textit{3D volumetric} samples (\textit{i.e.,} microfluidics, 3D cell models, and tissue sections). (4) This feature provides significant flexibility within the GenLFI framework, accommodating various optical setups and sample types. We conducted experiments using a full-frame CMOS camera and achieved a single-shot FOV exceeding \(500 \, mm^2\) while maintaining sub-pixel resolution. These advancements extend the boundaries of LFI technology in biomedical high-throughput \textit{dynamic} imaging tasks. 

\section*{Results} \label{sec2}
\subsection*{Imaging framework with unsupervised model}\label{result-Overview}
\begin{figure}[t]%
\centering
\includegraphics[width=1.0\textwidth]{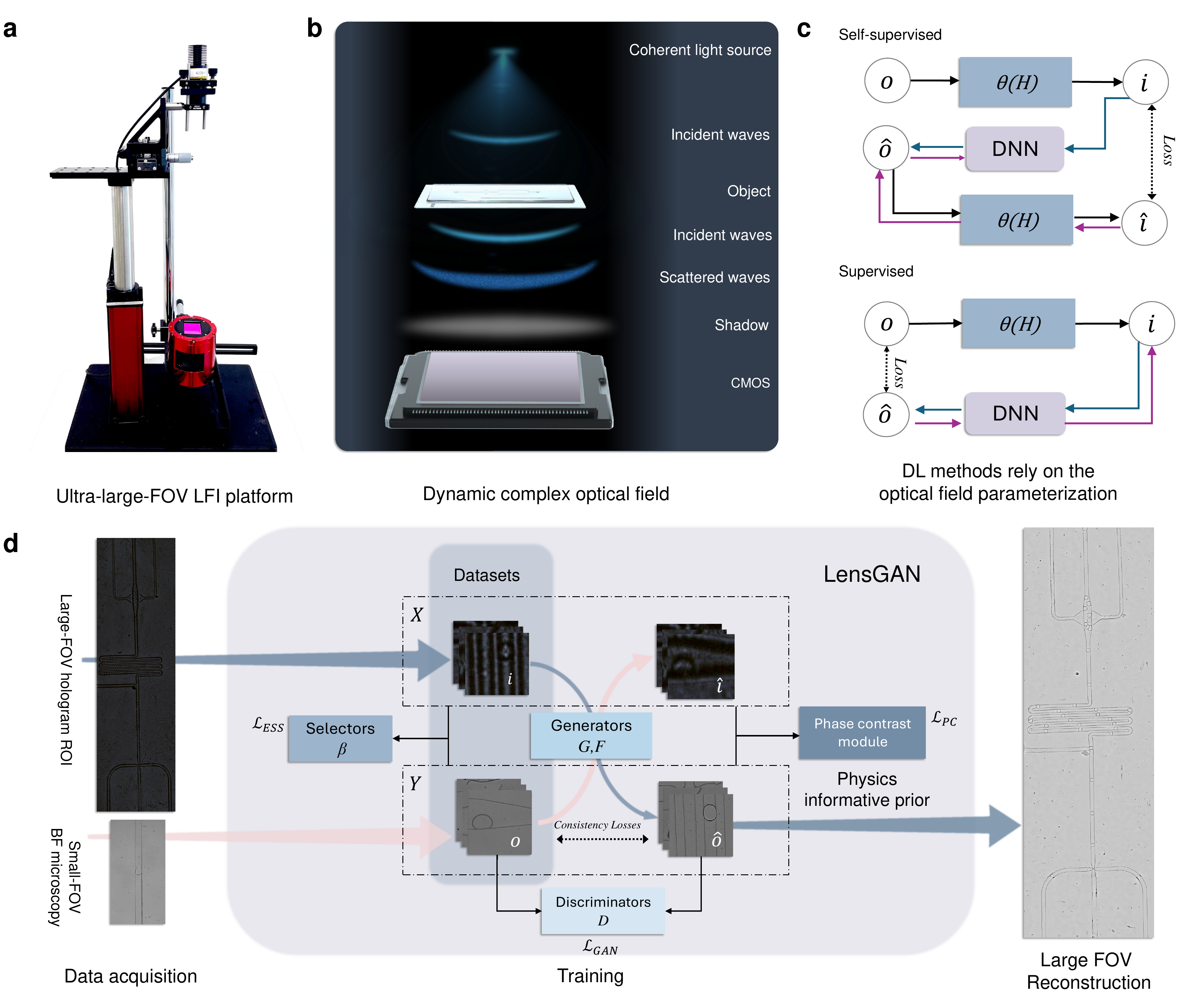}
\caption{Overview of GenLFI framework. \textbf{a}, Ultra-large FOV LFI setup with a telescope camera. \textbf{b}, Dynamic `complex' optical field: as a coherent light source illuminates the sample, scattered waves interfere with the undistributed incident wave, forming a complex hologram above the CMOS sensor through the interference pattern and shaded projection. \textbf{c}, Existing DL methods for LFI with the measured and reconstructed holograms (\(i\) and \(\hat{i}\)) and the bright-field images (\(o\) and \(\hat{o}\)): \textbf{top}, self-supervised deep neural network \cite{huangSelfsupervisedLearningHologram2023}; \textbf{bottom}, supervised deep neural networks. They both need a parameterized optical field as prior knowledge, denoted as \(\theta(H)\).  \textbf{d}, The workflow of GenLFI, which employs a physics informative prior instead of \(\theta(H)\) in \textbf{c}.}\label{fig-overview}
\end{figure}

Incorporating physics-based priors is essential in LFI. Previous LFI reconstruction methods are model-based and utilize a forward imaging model, \(H\), which integrates with the imaging system to relate the measured holograms \(i\) to the reconstructed image \(\hat{o}\) \cite{BarbastathisOnUseDLCI2019}.The LFI hologram reconstruction can thus be formulated as an inverse problem \cite{huangSelfsupervisedLearningHologram2023}: 
\begin{equation}
    \hat{o} = arg\min_{o}\mathcal{L}(H(o),i) + R(o)\label{eq-lfitask},
\end{equation} 
where \(i \in \mathbb{R_\text{i}}^{IK^2}\) denotes the measured holograms after vectorization; \(I\) represents the number of measured raw holograms, each with a dimension of \(K \times K\); \(o\) and \(\hat{o} \in \mathbb{R_\text{o}}^{K^2} \) are the real-world ground truth and predicted target image with a dimension of \(K \times K\) after vectorization, respectively, which in this study are bright-field images; \(\mathcal{L}(\cdot)\) and \(R(\cdot)\) are the loss function and regularization term, respectively. As shown in Fig.\ref{fig-overview}\textbf{c}, the parameterization of the optical field --- \(\theta(H)\), which inputs the optical setup parameters to decode the phase information, is essential for current LFI reconstruction algorithms in supervised \cite{rivensonPhaseRecoveryHolographic2018a, chenDeepLearningbasedHologram2023} and self-supervised \cite{huangSelfsupervisedLearningHologram2023} fashions. 

However, as discussed earlier, crafting an accurate physical forward (wave propagation) model \(H\) in (\ref{eq-lfitask}) within a complex optical field, as illustrated in Fig.\ref{fig-overview}\textbf{b}, presents significant challenges. The 3D sample, with vertical dimensions measured in micrometers (\(\mu m\)), can produce overlapped scattered waves from various axial distances and angles, particularly when interacting with mediums of different refractive indexes. In such scenarios, the optical field generates a complex hologram at the sensor, incorporating diffraction waves from different height planes and shadow projections. Furthermore, the optical field is subject to alterations caused by the movement of dynamic objects, which introduces additional distortions. These factors collectively contribute to difficulties in accurately describing, modeling, and reconstructing the optical field using current z-stack computational algorithms \cite{barkleyHolographicMicroscopyPython2018, chenDeepLearningbasedHologram2023, wangEHoloNetLearningbasedEndtoend2018, huangSelfsupervisedLearningHologram2023, jayapalaLensfreeHolographicImaging, chenEFINEnhancedFourier2023, zhangFocusNetClassifyingBetter2022, leeDeepLearningBased2023a, ozcanSingleShotAutofocusingMicroscopy2023}. A typical issue caused by 3D samples is the shadows causing information loss in the hologram: as shown in Fig.\ref{fig-compare}\textbf{a}, the z-stack backpropagating method for 2D LFI (Autofocus \cite{zhang3DImagingOptically2017, mckayLensFreeHolographic2023, greenbaumWidefieldComputationalImaging2014, scherrerHolographicReconstructionEnhancement2022}) failed to reconstruct the shaded hologram inputs and brings channel dislocation; as it struggles to parameterize (\(\theta(\cdot)\)) the optical field (\(H\)) in Fig.\ref{fig-overview}\textbf{c} accurately (for discussion on 3D LFI, see Supplementary Information 2).

To address these challenges, we introduce GenLFI, an LFI framework, that directly utilizes an unsupervised model trained on randomly captured data. It comprises a universal optical setup (see \hyperref[meth-optset]{Methods}), and a physics-informed generative deep-learning network, \textit{i.e.}, LensGAN. As shown in Fig.\ref{fig-overview}\textbf{d}, LensGAN introduces three key enhancements to establish these physical prior correlations and optimize the generative model: (1) Compared to current CycleGAN-based networks \cite{zhuUnpairedImagetoImageTranslation2017, scherrerHolographicReconstructionEnhancement2022,caoLabelfreeIntraoperativeHistology2023} in imaging, we apply a selector network \(\beta\) \cite{xieUnalignedImagetoImageTranslation2021}  to enable the transformation of unaligned images, effectively handling objects that are not aligned across different FOVs from randomly captured data. (2) We design a discriminator network \(D\) that integrates the attention mechanism \cite{vaswaniAttentionAllYou2017} between the downsampling convolutional layers to improve the processing of fine details in amplitude (\textit{e.g.,} illumination details). (3) We devise and incorporate a phase-contrast module to apply a physics-informed soft constraint to mimic the mechanism of the phase-contrast microscope (PCM) and dark-field microscope (DFM) (details in \hyperref[meth-algo]{Methods}). 

These improvements eliminate the need for optical field modeling of \(H\) in Eq. (\ref{eq-lfitask}) (\(\theta(H)\) in Fig.\ref{fig-overview}\textbf{c}),  offering flexibility in setup design and robustness against environmental disturbance. The data acquisition is simplified and does not require the hologram and microscopic data to be paired or aligned. 
The workflow of GenLFI is as follows: (1) An unpaired, unaligned dataset, randomly acquired from the LFI setup and a microscope, serves as the training set. (2) During training, LensGAN maps the measured and reconstructed holograms (\(i\) and \(\hat{i}\)) and bright-field images (\(o\) and \(\hat{o}\)) into two domains, \(X\) and \(Y\), respectively (see Fig.\ref{fig-overview}\textbf{d}). (3) The trained LensGAN model is then deployed to reconstruct hologram patches, synthesized into large FOV microscopic images.

\subsection*{Physics-informed prior and performance}\label{result-Physics}
We hypothesize that optical propagation in LFI is a latent nonlinear, multiscale physics model obscured by noise, presenting a complex and ill-posed inverse problem. Rather than directly constructing a forward model \(H\) as in Eq. (\ref{eq-lfitask}), phase information can be leveraged to enhance training and improve the estimation of real amplitude images from noisy holograms. 
Therefore, we integrate a \textit{phase contrast module} with the baseline CycleGAN, incorporating physical constraints and inductive learning bias to construct LensGAN as a PINN (details in \hyperref[meth-algo]{Methods}). This module captures the inherent phase correlations between holograms and bright-field images, enforcing physical consistency between the hologram's phase details (\textit{e.g.,} structures) and the bright-field image through soft physics penalty constraints and regularization. 

\begin{figure}[t]%
\centering
\includegraphics[width=1.0\textwidth, height=0.9\textheight, keepaspectratio]{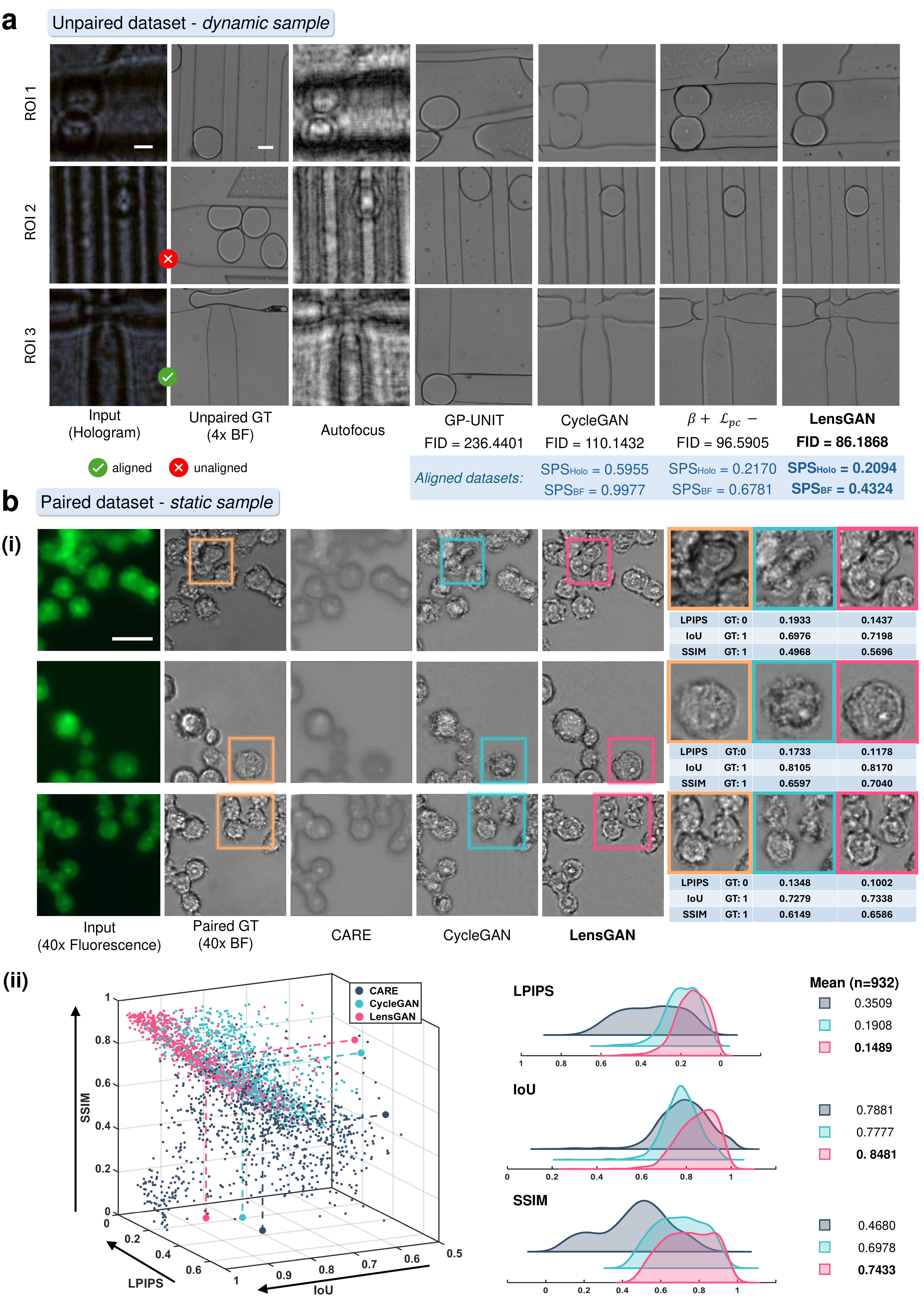}
\caption{Reconstruction performance on datasets, scale bar: \(100\,\mu m\). \textbf{a}, Unpaired datasets collected from microfluidics, trained in an unaligned manner. The entire unpaired dataset is evaluated using average FID. Additionally, we select the aligned unpaired ROI (i.e., the same region) and compute the average SPS score for the input hologram and the aligned BF ROI, respectively. \textbf{b}, Paired dataset of fluorescent/brightfield Caco-2 cells (\(40\times\)) \cite{triznaBrightfieldVsFluorescent2023}; \textbf{(i)} displays three ROIs from challenging (< mean) cases with detail illustrated in orange (GT), blue (CycleGAN \cite{zhuUnpairedImagetoImageTranslation2017}), and red (LensGAN), respectively. \textbf{(ii)} presents the quantitative results of the three methods across all 932 test sets.}\label{fig-compare}
\end{figure}

\textbf{Unpaired dataset:} Deep neural networks are effective at extracting semantic features in computer vision tasks but lack the physical constraints inherent to computational imaging, impacting reconstruction accuracy. To address this, a physics-informed prior can enhance imaging modality consistency, particularly for low-quality holograms. We first applied LensGAN upon our LFI setup with the \textit{unpaired and unaligned dataset}, \textit{randomly} captured from \textit{dynamic} microfluidics, as shown in Fig.\ref{fig-compare}\textbf{a}.  We employed the Fréchet Inception Distance (FID) metric \cite{heuselGANsTrainedTwo2017} to evaluate the entire test sets, where a lower FID indicates a better correlation between the distributions of generated images and ground truth. In addition, we selected aligned ROI from the dynamic dataset for further evaluation: We compute the Shape Preservation Score (SPS), a Wasserstein-distance-based metric over object-level geometric features, inspired by classical morphometric analysis \cite{lamThinningMethodologiesaComprehensive1992} and generative model evaluation \cite{arjovskyWassersteinGAN2017a}; Lower SPS means better shape alignment (see Supplementary Information 9).

In the complex optical field produced by dynamic 3D objects, LensGAN effectively reconstructed microfluidic channel structures from blurry, shadow-affected holograms in complex optical fields. As shown in ROI2 in Fig\ref{fig-compare}\textbf{a}, reconstruction from z-stack Autofocus \cite{hughesPyHoloscopePyHoloscopeDocumentation, mckayLensFreeHolographic2023, greenbaumWidefieldComputationalImaging2014, scherrerHolographicReconstructionEnhancement2022} shows the dislocation and inverted channel width affected by shadows from 3D objects. Ablation studies reveal that omitting the phase contrast loss (\(\mathcal{L}_\mathrm{PC} -\)) led to artifacts and hallucinated channels, influenced by shadow artifacts. In comparison to \textit{physics prior} guidance, we also reproduced the \textit{semantic generative prior} distillation approach in the state-of-the-art I2I algorithm GP-UNIT \cite{yangGPUNITGenerativePrior2023}. The result shows that results lost consistency despite their detailed image quality, verifying the semantic feature in holograms is difficult to extract. 

\textbf{Paired dataset:} To further quantify its cross-modality reconstruction performance, we also evaluated LensGAN on a paired dataset \cite{triznaBrightfieldVsFluorescent2023}, as shown in Fig.\ref{fig-compare}\textbf{b}. The fluorescent/bright-field cross-domain evaluation tests the models' generalizability across two imaging modalities. Specifically, we used learned perceptual image patch similarity (LPIPS) \cite{zhangUnreasonableEffectivenessDeep2018a}, binary mask intersection over union (IoU), and structural similarity index measure (SSIM) to assess reconstruction quality. We compared LensGAN with the supervised physics-informed model CARE \cite{weigertContentawareImageRestoration2018}, and the baseline unsupervised model CycleGAN \cite{zhuUnpairedImagetoImageTranslation2017}. While CARE preserves object structures, its image quality is insufficient for holographic reconstruction, indicating that current physics-informed image restoration models are inadequate for cross-modality holographic reconstruction. In addition, it is impractical to acquire paired datasets in LFI due to the requirements of light propagation through the samples. Conversely, CycleGAN achieves high perceptual quality in LPIPS but loses imaging modality consistency according to structural scores in IoU and SSIM.

\subsection*{FOV, resolution and reconstruction speed}\label{result-Charact}
\textbf{FOV}: In LFI, the achievable FOV captured in a single shot (one-shot FOV) is limited by the numerical aperture (NA) of the point light source and the sensor size \cite{greenbaumImagingLensesAchievements2012}, and the distance \(f\) between the light source and the detector, \textit{e.g.,} the CMOS sensor in Fig.\ref{fig-overview}\textbf{a}. While previous research \cite{zhang3DImagingOptically2017, chenEFINEnhancedFourier2023, leeDeepLearningBased2023a} has demonstrated some flexibility in adjusting this distance within a millimeter (\(mm\)) range, achieving the best lateral resolution for single hologram reconstruction requires a specific optimized distance. This optimized distance \cite{serabynResolutionOptimizationOffaxis2018} for resolution in a single coherent beam, applicable to both in-line and off-axis LFI, can be calculated as: 
\begin{equation}
    f_{opt}=\frac{1}{\frac{1}{d}-\frac{\lambda}{Np^2}},
    \label{eq-optdis}
\end{equation} 
where \(\lambda\) is the parameter wavelength, N is the number of pixels of the image sensor, \(p\) denotes the pixel size, and \(d\) is the distance between the object and sensor. The optimized distance is essential for generating high-resolution holograms, the cornerstone of existing LFI holographic reconstruction algorithms. Therefore, with a given NA, the FOV of existing LFI systems is constrained by the distance \(f \approx f_{opt}\). LensGAN overcomes this limitation by decoupling the reconstruction process with an optical setup, eliminating the need for precise diffraction pattern decoding. It can be validated by the effectiveness of the fluorescent dataset shown in Fig.\ref{fig-compare}\textbf{b}.  This enables the GenLFI framework to surpass the \(f_{opt}\) constraint, offering flexibility on the distance \(f\). The FOV limitation is then primarily governed by the sensor size. In our demonstration, we achieve a FOV of \(553 \, mm^2\) on a full-frame sensor and \(57\, mm^2\) on a \(1^"\) sensor (see Fig.\ref{fig-spheroids}\textbf{b} with the setup in \hyperref[meth-optset]{Methods}). Further improvements in FOV are possible through an upgraded optical configuration by using a higher NA light source and a larger and higher-resolution image sensor. 

\begin{figure}[t]%
\centering
\includegraphics[width=1.0\textwidth]{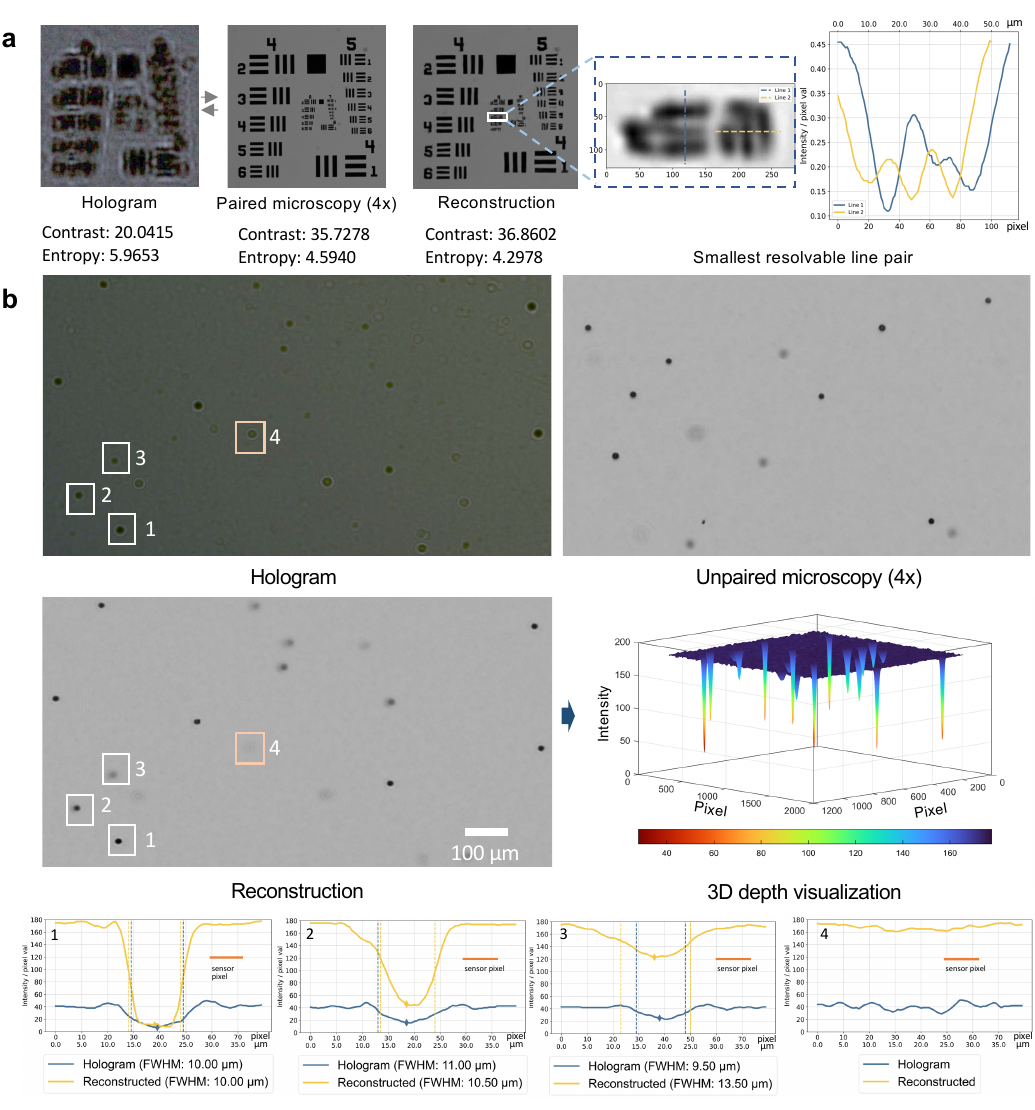}
\caption{Resolution test on \(5.9 \, \mu m / pixel\) CMOS sensor. \textbf{a}, Imaging of the USAF 1951 resolution test chart: the input hologram was aligned with the ROI of the smallest region of the test chart in a bright field. The white box highlights the reconstructed resolvable area, while the cross-sectional profile curves on the right demonstrate the distinguishable line pairs. \textbf{b}, \(10 \, \mu m\) diameter beads imaging. The beads, which were black and opaque, were reconstructed by LensGAN using their features from small dirt and transparent droplets. We select 3 true-positive (\textbf{1-3}) and 1 true-negative object (\textbf{4}) at different depths and visualize them by cross-sectional profiles. In this particular case, the 3D depth map can be visualized based on pixel intensities from the reconstruction. }\label{fig-resolution}
\end{figure}


\textbf{Resolution}: The image resolution attainable by GenLFI is affected by two physical parameters. First, the Nyquist-Shannon sampling theorem dictates that the pixel size of the CMOS sensor inherently constrains the spatial resolution of holograms. This limitation suggests that details finer than the sensor's pixel pitch cannot be adequately captured in the hologram \cite{alma9993244227901488}. Notably, the real-time imaging necessitates single-shot capture, and the pixel-shifting super-resolution strategy \cite{greenbaumMasklessImagingDense2012, greenbaumFieldportableWidefieldMicroscopy2012, zhang3DImagingOptically2017} is not feasible. Second, the reconstructed image quality is significantly influenced by the dimension of the input image of the model (see \hyperref[meth-training]{Methods}). In particular, LensGAN's training is memory-intensive on GPU due to the need to load six networks (see \hyperref[meth-algo]{Methods}), limiting the input image dimensions due to hardware constraints. 

On the full-frame \(5.9 \, \mu m/px\) sensor,  we imaged a 1951 USAF resolution test chart and conducted a cross-sectional profile analysis following the standard routine \cite{luoPixelSuperresolutionUsing2016, luoPixelSuperresolutionLensfree2019, kimPtychographicLenslessBirefringence2023}. We selected an area within the measured hologram where the smallest line bar is not resolvable (see Fig.\ref{fig-resolution}\textbf{a}). After reconstruction, we found that the smallest resolvable line pair was element 4 of group 6 (Fig.\ref{fig-resolution}\textbf{a}), corresponding to the best observing lateral resolution \(5.52\, \mu m\), which falls below the sensor’s pixel size \(5.94 \, \mu m\). Furthermore, the evaluation metrics, \textit{i.e.,} the FID and SSIM, are 22.6855 and 0.8203, respectively, when comparing the reconstruction with the ground truth obtained by a bright-field microscope. The standard deviation contrast is enhanced from 20.0415 to 36.8602; the image entropy decreased from 5.9653 down to 4.2978. Notably, using a higher-resolution sensor is a straightforward approach to improve lateral resolution. On a \(3.45 \, \mu m/px\) sensor, the resolution is \(3.11 \, \mu m/px\).

We hypothesize that the diversity and quantity of samples in the training dataset also influence the final resolvable resolution. In the 1951 USAF resolution test, the observed object is only one test chart (see \hyperref[meth-usaf]{Methods}), which will likely cause the over-fitting problem. To address this, we conducted another experiment with \(10 \, \mu m\) diameter beads, increasing data diversity and quantity. The reconstructions clearly recover and resolve the bead shapes from different regions (see Fig.\ref{fig-resolution}\textbf{b}). Unlike traditional LFI confined to a thin focal plane, our results demonstrate a higher depth of field. This allows us to distinguish objects at various depths within the water flow (see \hyperref[meth-beads]{Methods}). The full-width-at-half-maximum (FWHM) shows that in cases 1 and 2, the reconstruction for focused beads keeps positional consistencies with FWHM difference within \(0.5 \, \mu m\) on a \(5.9 \, \mu m/px\) sensor. Case 3 shows the effect of shadow, which is a feature learned from microscopy. Case 4 shows the true negative of perturbation of non-opaque hologram fringe, which is formed by bead size smaller than \(10 \, \mu m\) \cite{isikmanLensfreeOnChipMicroscopy2011}. We visualize the cross-sectional profiles of three beads at different depths in Fig.\ref{fig-resolution}\textbf{b}, along with their depth information encoded by pixel intensity. This capability for depth recovery paves the way for potential 3D high-throughput imaging using GenLFI.

\textbf{Speed}: Previous studies \cite{huangSelfsupervisedLearningHologram2023, alma9993244227901488} have suggested that DL can expedite the reconstruction process through a pure end-to-end approach. Compared to previous DL approaches, LensGAN's capability for dynamic imaging is enhanced by eliminating the need for multiple (\(I\) number) hologram measurements, sub-pixel super-resolution, and Fourier phase and amplitude pre-processing. Additionally, its fast model reference speed further contributes to its efficiency. Once deployed, the trained model relies only on a single network (\textit{i.e.,} the generator \(G\) built by a 9-layer ResNet \cite{heDeepResidualLearning2016a}) for direct reconstruction from the input. We evaluated LensGAN's performance (see \hyperref[meth-training]{Methods}) using \(512 \times 512\) pixels (\(px\)) hologram images. LensGAN was compared against the self-supervised model GedankenNet \cite{huangSelfsupervisedLearningHologram2023}, the supervised model U-Net \cite{unet}, and the iterative algorithm MHPR \cite{rivensonPhaseRecoveryHolographic2018a}, as outlined in Table \ref{tab1}. LensGAN achieved an equivalent inference time of \(0.0031 \, {s \cdot mm}^{-2}\) for a \(1 \, mm^2\) area of our microfluidic chip (see Fig.\ref{fig-microfluidics}), showing its exceptional speed and efficiency in hologram reconstruction. Note also that efficient model compression techniques for GANs \cite{2021OnlineCompressGAN} can further reduce the computational cost by \(46 \times\), but we maintain the original model size to avoid resolution degradation.

\begin{table}[h]
    \centering
    \caption{\centering Holographic reconstruction speed (for \(I\) input raw hologram(s) in \(512^2\, px\))}
    \label{tab1}
    \begin{tabular}{@{}cccccc@{}}
        \toprule
        Method & \makecell{LensGAN (\(I\)=1)} & \makecell{GedankenNet (\(I\)=2)} & \makecell{U-Net (\(I\)=2)} & \makecell{MHPR (\(I\)=2)} \\
        \midrule
        Reconstruction time & \textbf{0.0015\(s\)} & 0.05\(s\) & 0.02\(s\) & 0.98\(s\) \\
        Inference time  & \textbf{0.0015\(s\)} & 0.0032\(s\) & 0.0013\(s\) & 0.0625\(s\) \\
        \#Parameters & 33.8M & 39.4M & 31.0M & - \\
        \bottomrule
    \end{tabular}
    \footnotetext{* M: million. Reconstruction time refers to the total duration from raw hologram input through phase retrieval to model inference. Inference time is measured from the retrieved phase input to the model output, excluding the phase retrieval step (i.e., Fourier transform). All evaluations were conducted on a single PC with an i9-13900K CPU and an RTX 3090 GPU. Note that GedankenNet and MHPR require multiple measurements for input data, and U-Net requires manual annotation for training.}
\end{table}

\subsection*{Spheroids monitoring} \label{result-Spheroids}
\begin{figure}[t]
\centering
\resizebox{\textwidth}{!}{%
    \includegraphics{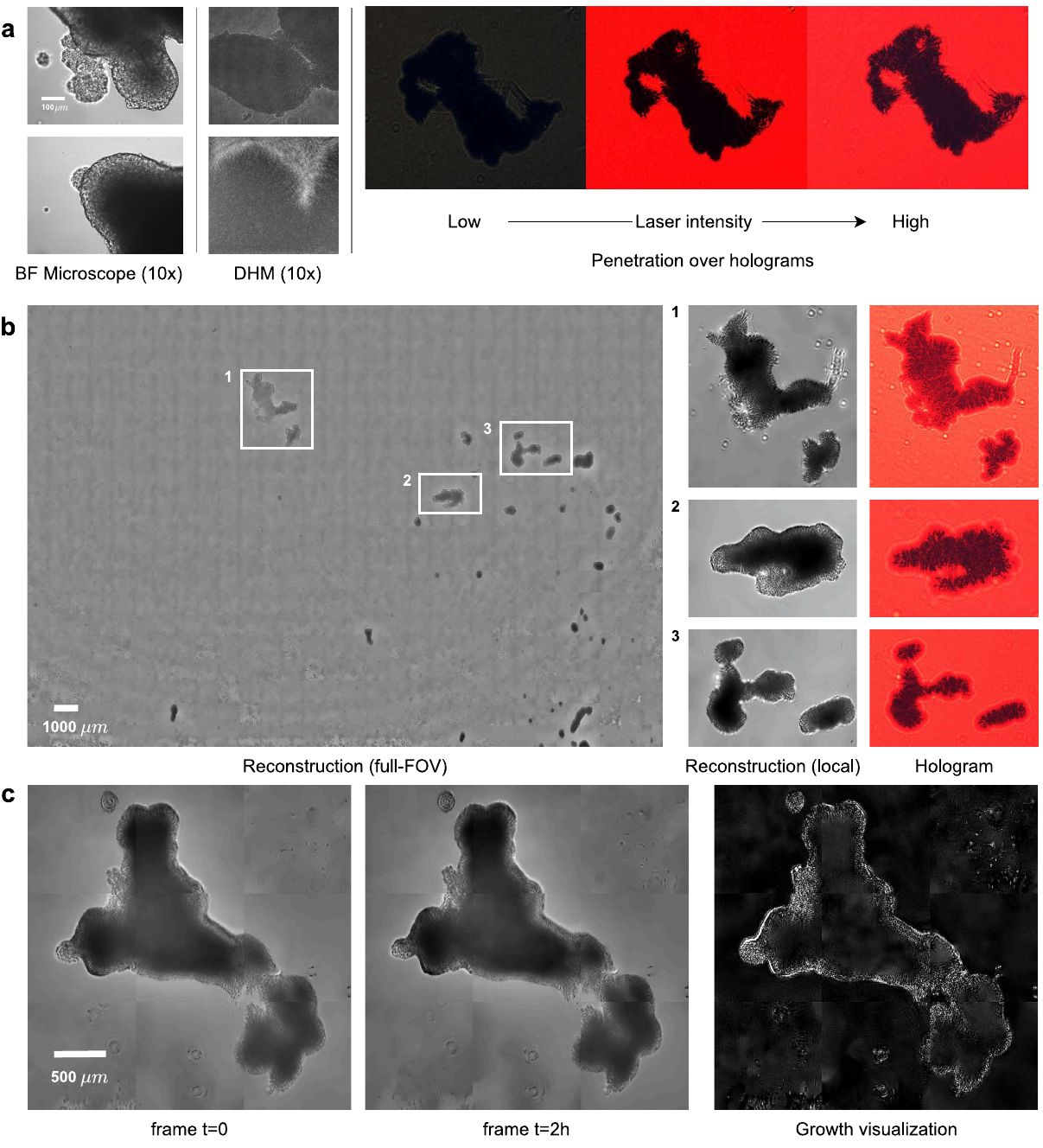}
}
\caption{
Reconstruction of unpaired cell spheroids. \textbf{a}, MeT-5A NF2KO spheroid images acquired using different imaging systems: bright-field microscopy with a 10x objective lens (Microscope), digital holographic microscopy (DHM) with a 10x objective lens, our GenLFI platform (LensGAN). With the laser's high intensity, deeper structural information of the 3D spheroid is penetrated and illustrated. \textbf{b}, Full-FOV reconstruction by LensGAN on an unseen well, post-processed by global panorama stitching algorithm (see Supplementary Information 5) and local image-stitching \cite{weberAutomatischeZusammenfuehrungZertrennter2022}. The structures are readily resolvable in the penetrated light parts at the edges of spheroids. \textbf{c}, Swelling spheroids were observed over a 2-hour time-lapse, the maximum duration allowable outside the incubator. The growth region is highlighted in the bright areas of the differential image between the two frames.
}
\label{fig-spheroids}
\end{figure}
Cell spheroids, as straightforward yet extensively utilized multi-cellular 3D models, emerge from the natural inclination of most adherent cells to cluster together \cite{3DSpheroids3D}. The imaging of spheroids poses significant challenges due to their considerable thickness, which can reach up to hundreds of microns, and their high density. As noted previously, conventional holographic imaging systems, including LFI systems, involve complex optical fields (Fig.\ref{fig-overview}\textbf{a}), resulting in contaminated holograms with overlapping scattered patterns and shadows. In this case, both lens-based digital holographic microscopy (DHM) and the Autofocus algorithm \cite{zhang3DImagingOptically2017, mckayLensFreeHolographic2023, ozcanSingleShotAutofocusingMicroscopy2023} cannot reconstruct the 3D feature of spheroids (see Fig.\ref{fig-spheroids}\textbf{a}).

With the addition of a light contrast-enhancing polarization filter, a bright-field microscope equipped with a 10x objective lens can discern a partial sectional structure at the periphery (Fig.\ref{fig-spheroids}\textbf{a}, Microscope (10x)). A comparable light penetration capability can also be achieved using our LFI platform by employing a red laser light source with a wavelength of \(635 \, nm\). Higher laser intensity allows for the extraction of more 3D features from holograms. The results indicate that LensGAN can not only reconstruct 3D structures but also maintain robustness under varying light conditions. This demonstrates a level of resilience to environmental factors that current LFI methods do not achieve.

Fig.\ref{fig-spheroids}\textbf{b} shows that our approach can reconstruct the spheroid edge structure and shapes from the measured holograms in a large FOV of an engineered cancer model. Notably, spheroids exhibit lateral structures that might replicate some of the more complex cellular environments, thereby mimicking more relevant scenarios while using high throughput platforms. Recent research \cite{bulcaenPrimeEditingFunctionally2024} has shown that time-lapse label-free monitoring is an effective and reliable evaluation tool for evaluating the effects of gene editing on organoids. Following this strategy, we recorded 2-hour time-lapse holograms and reconstructed them using LensGAN. The frame difference highlights the growth distribution on the spheroid, which has not been shown in such a large FOV in real-time before. Leveraging the advantages of the large FOV, our approach proves effective for monitoring the multiparametric spheroid growth patterns, which will likely allow for insights into biomechanical features, cell-cell interactions, and readily quantifiable multiparametric features of spheroid outgrowth assays in a scalable and cost-effective manner.

\subsection*{Microfluidics screening} \label{result-Micofl}

\begin{figure}[t]%
\centering
\includegraphics[width=1.0\textwidth]{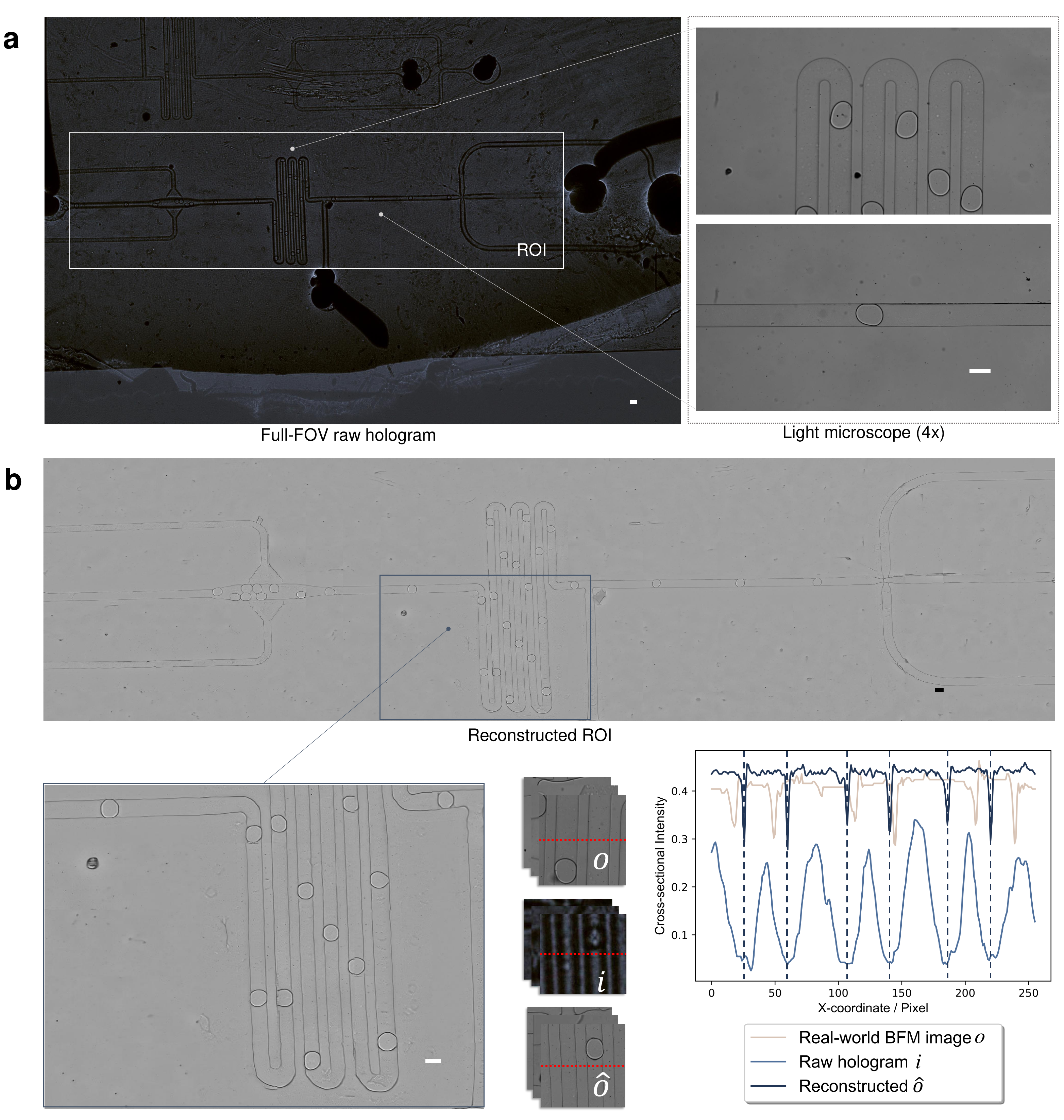}
\caption{Reconstruction of a microfluidic chip (\(30\times5 \, mm\)), scale bar: \(200 \, \mu m\). \textbf{a}, ROI in the hologram, \textit{i.e.,} the functional region containing dynamic droplets. \textbf{b}, Synthesized reconstruction result of the ROI (also see Supplementary video 1). We also analyzed the cross-sectional profiles at the center of 3 unpaired patches, labeled \(i\), \(o\), and \(\hat{o}\), as denoted by the red dash lines in the middle. }\label{fig-microfluidics}
\end{figure}

Microfluidic platforms have transformed biomedical research by enabling high-throughput experiments with biological samples down to the scale of single cells or molecules \cite{nanDevelopmentFutureDroplet2024}. Despite their increasing adoption, a critical challenge remains. Existing high-resolution imaging techniques cannot simultaneously monitor multiple ROIs on a single chip. For instance, imaging the formation of droplets in droplet-based microfluidics often impedes simultaneous observation of downstream operations, such as droplet sorting or merging, occurring in a different ROI \cite{nanDevelopmentFutureDroplet2024}. Most current LFI methods require multiple measurements from different positions and are unable to capture real-time microfluidic droplets. The only viable approach is forward optical modeling. However, this approach is impractical for dynamic objects that change shape.

Our GenLFI approach overcomes these limitations. We demonstrate its capabilities in a microfluidic experiment, generating and dynamically imaging droplets across various chip sections concurrently. Data collection for the training dataset of this microfluidic experiment (see \hyperref[meth-micfl]{Methods}) took approximately 20 minutes. Unlike most LFI systems, this procedure is automated on random ROIs, requiring no manual annotation. We acquired two-domain training data using our LFI system and a conventional bright-field microscope, labeled as \(i\) and \(o\), respectively  (see Fig.\ref{fig-overview}\textbf{b}). These two-domain images, \(i\) and \(o\), constituted the `GenLFI-Microfluidics' dataset for LensGAN training and testing. Reconstruction results (see Fig.\ref{fig-microfluidics})  demonstrate that the trained LensGAN can generate high-resolution (\(1792 \times 768 \, px\)) bright-field images (\(\hat{o}\)) across the entire ROI, covering a full microfluidic chip.

The results show that our approach not only preserves the global structure from the original hologram to the generated image but also ensures consistent transverse x-coordinate positioning of features, as evident in the re-focusing process shown in Fig.\ref{fig-microfluidics}\textbf{b}. Furthermore, the synthesized large FOV image (Fig.\ref{fig-microfluidics}\textbf{b}) reveals that although the trained LensGAN model infers each patch independently (see \hyperref[meth-micfl]{Methods}), the microfluidic channels are aligned consistently across patches. This suggests the robustness of our approach in handling large-scale reconstruction tasks. This capability can facilitate monitoring and visual-based automation for intricate microfluidic systems.

\section*{Discussion} \label{sec3}
GenLFI, a generative physics-informed deep learning framework for LFI, eliminates the need for traditional forward imaging models by introducing random training data through unsupervised learning and physics constraints and regularization. This framework allows imaging in dynamic, noisy optical fields, which was previously unattainable with model-based LFI using model-based methods. For 3D samples, in a single shot with sub-pixel resolution, GenLFI achieves an over 20 times larger FOV than current LFI systems \cite{imecLensfreeImagingCompacta, alma9993244227901488}, enabling real-time monitoring for in vivo (\textit{e.g.,} zebrafish, see Fig.S1 in Supplementary Information) or in vitro (\textit{e.g.,} cell-in-droplets, 3D cell models) studies.  GenFLI is compatible with higher-resolution sensors and multi-angle illumination for potential 3D LFI capabilities. Images are reconstructed using a DL network, significantly faster than traditional iterative techniques \cite{barkleyHolographicMicroscopyPython2018,castanedaPyDHMPythonLibrary2022,berdeuReconstructionInlineHolograms2019,denisInlineHologramReconstruction2009,rivensonPhaseRecoveryHolographic2018a}, without requiring multi-position measurements and data pre-processing. These advancements demonstrate the potential of leveraging generative AI to overcome optical limitations. 
 
Our investigation into the physical correlation between holograms and bright-field images achieved leading results based on various metrics. However, challenges remain with such I2I-based approaches, primarily concerning data quality, quantity, and diversity. Training GANs can be challenging due to potential issues like model collapse and non-convergence. While exploring diffusion models \cite{hoDenoisingDiffusionProbabilistic2020,zhaoEGSDEUnpairedImagetoImage2022,sasakiUNITDDPMUNpairedImage2021,liuSBImagetoImageSchr2023} as an alternative shows promise, their limitations in handling unpaired and unaligned image translation tasks require further investigation. 

It is notable that, rather than a low-cost replacement for traditional lens-based optical systems, we consider LFI systems as high-content imaging tools for data-driven biomedical applications. For instance, a modest FDA-approved drug library can contain around 2000 samples and 2 million or more pairwise combinations \cite{kulesaCombinatorialDrugDiscovery2018}; GenLFI could utilize the repetitive nature of these samples’ large-scale high-content data. As GANs lack generalization abilities, LensGAN, a very initial exploration under the GenLFI framework, is more suitable for large-scale screening automation on similar modality samples. In the future, the combination of GenLFI and foundation large vision models (LVM) \cite{cuiLargeScaleFineGrained2018,baiSequentialModelingEnables2023} holds promise for achieving a more generalized reconstruction capability for various samples and light field conditions. In addition, leveraging chips with advanced and specified architectures \cite{jiaDissectingGraphcoreIPU2019,arcelinComparisonGraphcoreIPUs2021,pengEvaluatingEmergingAI2023} and cloud platforms could enable real-time image reconstruction with these large models. 

Despite limitations due to optical conditions, GenLFI's high-throughput capability opens doors for various biomedical applications. GenLFI offers a large FOV, enabling rapid analysis of large biological samples with minimal equipment. 
Furthermore, GenLFI's cost-effective and compact setup facilitates routine monitoring of cell culture and spheroid growth directly within incubators, reducing phytotoxicity while enabling continuous monitoring of dynamic cellular processes across a large FOV.  The testing of drug libraries based on \textit{dynamic} processes, utilizing single-cell/cell models or droplet-based microfluidic systems, can now be monitored \textit{in parallel}, paving the way for developing next-generation, autonomous, high-throughput drug screening systems.

\section*{Methods} \label{sec4}

\subsection*{Optical setup} \label{meth-optset}
GenLFI utilizes a versatile optical setup for capturing holograms. The optical setup comprises a light source, a CMOS sensor, and a test bench to stabilize their positions. We employed two types of light sources, including a blue LED light source for enhanced resolution (used for \hyperref[meth-micfl]{microfluidics}, \hyperref[meth-usaf]{test chart}, \hyperref[meth-beads]{beads}) and a red laser light source for deeper penetration depth (used for \hyperref[meth-sphimg]{spheroids}). For LED illumination, the light source is Thorlabs M420L3  (wavelength \(420 \, nm\)) blue LED, with power supply (15V 1A) and T-Cube LED Driver (Thorlabs LEDD1B). For laser illumination, we used an optical fibre-coupled red laser source (Thorlabs S1FC635, wavelength  \(635 \, nm\)). The mechanical setup consists of a cage plate (Thorlabs CP33/M), a mounting bracket (ThorLabs 973/579-7227), a \(25 \, mm\) linear stage (Thorlabs XR25C) with an end-mounted micrometer, a post holder (Thorlabs LPH200), a post (Thorlabs P350/M), and an adjustable platform (Thorlabs C1519). The telescope camera (ZWO ASI 2400MC Pro) features a full-frame Sony IMX410 CMOS sensor, \(36\times24 \, mm\) in size, with a  \(5.94 \, \mu m/px\) resolution. Another camera used for tissue imaging is Thorlabs CS895MU, with 
\(14 \times 7 \, mm\) size and \(3.45 \, \mu m/px\) resolution. Mounted on an experimental bracket (FAMA MV test stand), the camera aligns with the light source and the imaging object along a single axis. The distance from the light source to the object’s bottom plane is \( 430 \, mm\), and from the object’s bottom plane to the CMOS sensor is \(12.5 \, mm\).

\subsection*{LensGAN architecture} \label{meth-algo}

LensGAN, a core component of GenLFI, reconstructs high-resolution images from captured holograms. LensGAN generally considers the measured and reconstructed holograms (\(i\) and \(\hat{i}\)) and the measured and generated bright-field images (\(o\) and \(\hat{o}\)) into two domains, \(X\) and \(Y\), respectively. It leverages a GAN framework with a cycle-consistent training strategy to learn the complex mapping between holograms (domain $X$) and corresponding bright-field images (domain $Y$). The architecture of LensGAN comprises 6 networks: generators of reconstruction and reverse mappings \(G:X \rightarrow Y\) and \(F:Y \rightarrow X\), selectors  \(\beta_X\) and \(\beta_Y\), and discriminators \(D_X\) and \(D_Y\). During training, all 6 networks function following the training loop as shown in Fig.\ref{fig-overview}\textbf{d} and supplementary Fig.~S3. In the testing phase, only the trained generator \(G\) is used for reconstruction. The outputs of LensGAN are reconstructed bright-field images \(\hat{o} = G(i)\) and generated holograms \(\hat{i} = F(o)\).

\subsubsection*{Neural networks}
\textbf{Generators} \(G\) and \(F\) handle the fundamental image transformation task, while the remaining 4 networks contribute to the training process. \(G\) and \(F\), each composed of 9 residual blocks (ResNet \cite{heDeepResidualLearning2016a}), feature an image encoder for extracting image characteristics and a decoder for generating images. The \textbf{selector} network \(\beta\) \cite{xieUnalignedImagetoImageTranslation2021} is responsible for pairing images based on feature alignments and is constructed with an adaptive average pooling layer, 4 convolutional networks, and one fully connected network (FCN) with a Softmax layer. The \textbf{discriminator} \(D\) is designed to differentiate between images, emphasizing structural information and intricate details within the complex amplitude. Recent advancements in Transformers \cite{vaswaniAttentionAllYou2017} have demonstrated the equivalence of multi-layer and one-layer attention mechanisms \cite{wuSGFormerSimplifyingEmpowering2024}. By leveraging this finding, we designed \(D\) with a Markovian architecture, merging a one-layer attention module with 5 downsampling convolutional layers to augment amplitude details while maintaining minimal computational overhead (see Supplementary Information 4).

\subsubsection*{Loss functions}
Combining these networks, the training loop adheres to the cycle structure \cite{zhuUnpairedImagetoImageTranslation2017}, guided by several loss functions.

\textbf{Phase contrast loss:}
Instead of directly using the Fourier transform (FT) to pre-process the holograms and decode the phase in MHPR \cite{rivensonPhaseRecoveryHolographic2018a, huangSelfsupervisedLearningHologram2023},  we utilize the phase information as a soft learning bias during the learning process.

To augment phase structural feature information during the training, we devised a \textit{phase contrast module} positioned to constrain and regularize the physical correlation between the hologram (\(X\)) and bright field (\(Y\)) domains. In LFI the observed samples are typically transparent or semi-transparent, resulting in holograms that encode phase variations rather than intensity-based structural features. These holograms are typically captured using coherent, single-wavelength illumination, which lacks intrinsic contrast and leads to duotone or low-contrast interference patterns. 

Phase contrast microscopy \cite{zernikePhaseContrastNew1942} enhances contrast by converting phase shifts from refractive index variations into detectable intensity differences. This is achieved using a phase ring in the objective’s back focal plane to shift the background light's phase, producing interference that reveals fine structural details, especially in transparent biological samples.

Inspired by this technique, we design a computational analog that introduces a soft inductive bias into the learning process.  Specifically, our model includes a \textit{phase-contrast loss} \(\mathcal{L}_\mathrm{PC}\) that encourages the generator to preserve and translate frequency-domain phase structures between the hologram and bright-field domains. By applying a multi-scale structural similarity index measure (MS-SSIM \cite{wangMultiscaleStructuralSimilarity2003}) on the amplitude spectra of the Fourier-transformed images, we promote consistency in the spatial frequency content and thereby encourage the model to learn phase-sensitive transformations: 
\begin{equation}
    \min_{G}\mathcal{L}_\mathrm{PC}(X,Y) = \mathbb{E}_{i \sim P_X}\left[1 - \operatorname{Sim}(|\mathcal{F}(i)|,|\mathcal{F}(\hat{o})|)\right]
\end{equation}
\label{eq-Lpc}.
Here, \(|\mathcal{F}(\cdot)|\) denotes the magnitudes of the individual frequency components obtained from the image's Fourier transform (\(\mathcal{F}\)), and \(P_{X}\) is the probability distribution for \(i\).  \(\operatorname{Sim}\) is the MS-SSIM function applied to the amplitude spectra, defined as:
\begin{equation}
    \operatorname{Sim}(x,y) = \left[l_M(x,y)\right]^{\gamma_M} \cdot \prod_{j=1}^M \left[c_j(x,y)\right]^{\delta_j} \left[s_j(x,y)\right]^{\epsilon_j}\label{eq-msssim},
\end{equation} with the terms
\begin{equation}
\begin{aligned}
        l(x,y) = \frac{2\mu_x\mu_y + C_1}{\mu_x^2 + \mu_y^2 + C_1}, \\ c(x,y) = \frac{2\sigma_x \sigma_y + C_2}{\sigma_x^2 + \sigma_y^2 + C_2}, \\ s(x,y) = \frac{\sigma_{xy} + C_3}{\sigma_x \sigma_y + C_3} \label{eq-ssimterm},
\end{aligned}
\end{equation}
where \(x\) and \(y\) are two input images, \(M = 5\) denotes the highest scale number set for the system to iteratively apply a low-pass filter between \(x\) and \(y\).   \(\mu\) and \(\sigma^2\) stand for mean value and variance, respectively, implemented by 2D convolutions. \(l(x,y)\), \(c(x,y)\), and \(s(x,y)\) represent luminance, contrast, and structure comparison measures between \(x\) and \(y\), with control weights \(\delta_1 = \epsilon_1 = 0.0448\),  \(\delta_2 = \epsilon_2 = 0.2856\), \(\delta_3 = \epsilon_3 = 0.3001\), \(\delta_4 = \epsilon_4 = 0.2363\), and \(\gamma_5 = \delta_5 = \epsilon_5 = 0.1333\). LensGAN enforces all inputs as 8-bit images with a pixel value range of 0 to 255; the constants are set as \(C_1 = (0.01\times255)^2\),   \(C_2 = (0.03\times255)^2\), and \(C_3 = C_2 / 2\). 

This loss not only guides the model to preserve fine structural details that are typically encoded in phase, but also helps prevent unrealistic geometric distortions across domains (see Fig.\ref{fig-compare}a) by enforcing fidelity in high-frequency phase content (see the FRC evaluation detailed in Supplementary Information 10).


\textbf{Effective sample size loss:}
we adopted the sample reweighting strategy for selecting the aligned image pairs \cite{grettonCovariateShiftKernela, xieUnalignedImagetoImageTranslation2021}.
Note that in Eq. (\ref{eq-Lpc}), we added \(\mathcal{L}_\mathrm{PC}\) only for the mapping \(G:X \rightarrow Y\), the transformation process of holographic reconstruction, with a control weight \(\tau_\mathrm{PC}\). Similarly,  only for \(G\), the selector network \(\beta_X\) and \(\beta_Y\) is regularized by the effective sample size loss \cite{xieUnalignedImagetoImageTranslation2021}:

\begin{equation}
    \mathcal{L}_{\mathrm{ESS}}(X, Y) = \|\beta_X\|_2 + \|\beta_Y\|_2, \quad 
    \min_{\beta_X, \beta_Y} \mathcal{L}_{\mathrm{ESS}}(X, Y). \label{eq-beta}
\end{equation}

where \(\|\cdot\|_2\) denotes the L2 norm.

Selectors \(\beta_X\) and \(\beta_Y\) take in real images (i.e., hologram \(i\) and bright field \(o\)) and produce importance weights \(\beta_X(i)\) and \(\beta_Y(o)\) for each sample. These weights are used to scale the unweighted generator loss components (e.g., cycle-consistency, adversarial loss) dynamically.

\textbf{Semantic consistency losses:}
For the whole cycle (reconstruction mapping \(G: X \rightarrow Y\) and reverse mapping \(F: Y \rightarrow X\)), there is a GAN loss function \(\mathcal{L}_{\mathrm{GAN}}\) to constrain \(G\), \(F\) and \(D\), with two consistency loss functions \(\mathcal{L}_{\mathrm{cyc}}\) and \( \mathcal{L}_{\mathrm{idt}} \) to further regularize \(G\) and \(F\) \cite{maoLeastSquaresGenerative2017, zhuUnpairedImagetoImageTranslation2017, xieUnalignedImagetoImageTranslation2021}. They are reweighted least squares adversarial loss to examine the generator:
\begin{equation}
    \begin{aligned}
    \min _{G, \beta_X} \max _{D_Y} & \mathcal{L}_{\mathrm{GAN}}(X, Y) \\ = & \mathbb{E}_{i \sim P_X} \beta_X(i)\left[\left(1-D_Y(\hat{o}\right)^2\right]+ \\ & \mathbb{E}_{o \sim P_Y} \beta_Y(o)\left[D_Y(o)^2\right],\label{eq-Lgan}
    \end{aligned}
\end{equation}
reweighted cycle consistency loss:
\begin{equation}
    \begin{aligned}
    \min_{G,F} \mathcal{L}_{\mathrm{cyc}}(X,Y) = \mathbb{E}_{i \sim P_X} \beta_X(i) \|i-F(\hat{o})\|_1  \\ +  \mathbb{E}_{o \sim P_Y} \beta_Y(o) \|o-G(\hat{i})\|_1 \label{eq-Lcyc},
    \end{aligned}
\end{equation}
and reweighted identity loss to encourage the model more conservative \cite{WhatMeaningIdentity}:
\begin{equation}
    \begin{aligned}
    \min_{G,F} \mathcal{L}_{\mathrm{idt}}(X,Y) = \mathbb{E}_{i \sim P_X} \beta_X(i) \|i-F(i)\|_1  \\ +  \mathbb{E}_{o \sim P_Y} \beta_Y(o) \|o-G(o)\|_1 \label{eq-Lidt},
    \end{aligned}
\end{equation}
where \(P_{X}\) and \(P_{Y}\) are the probability distributions for \(i\) and \(o\), respectively, and \(\|\cdot\|_1\) denotes the L1 norm.

The LensGAN's \textbf{full objective of loss functions} is combined with:
\begin{equation}
    \begin{gathered} \mathcal{L}_{main} = \min _{G, F} \max _{D_X, D_Y}  \mathcal{L}_{\mathrm{GAN}}(X, Y)+\mathcal{L}_{\mathrm{GAN}}(Y, X) \\
\quad+\tau_{\mathrm{cyc}} \mathcal{L}_{\mathrm{cyc}}(X, Y)+\tau_{\mathrm{PC}}\mathcal{L}_{\mathrm{PC}}(X,Y)+\tau_{\mathrm{idt}} \mathcal{L}_{\mathrm{idt}}(X, Y)  , \\  \mathcal{L}_{\beta_{X}} = \min _{\beta_X}  \mathcal{L}_{\mathrm{GAN}}(X, Y)+\tau_{\mathrm{ESS}} \mathcal{L}_{\mathrm{ESS}}(Y) , \\  \mathcal{L}_{\beta_{Y}}=\min _{\beta_Y}  \mathcal{L}_{\mathrm{GAN}}(Y, X)+\tau_{\mathrm{ESS}} \mathcal{L}_{\mathrm{ESS}}(Y)  ,\end{gathered} \label{eq-Lfull}
\end{equation}
where \(\mathcal{L}_{main}\) is used to train networks \(G\), \(F\), and \(D\), \(\mathcal{L}_{\beta_X}\) and  \(\mathcal{L}_{\beta_Y}\) are used to train \(\beta_X\) and \(\beta_Y\), respectively; the weights \(\tau_{\mathrm{cyc}}\), \(\tau_{\mathrm{PC}}\), \(\tau_{\mathrm{idt}}\), and \(\tau_{\mathrm{ESS}}\) control the relative importance of different losses and can be adjusted as hyperparameters. 

\subsection*{Training, testing and post-processing} \label{meth-training}
The training and testing follow a standardized CycleGAN PyTorch routine \cite{zhuJunyanzCycleGAN2025}, with details indicated in the supplementary code documentation. For all experiments, we set hyperparameters \(\tau_{\mathrm{cyc}} = 10\), \(\tau_{\mathrm{PC}} = 8\), and \(\tau_{\mathrm{ESS}} = \tau_{\mathrm{idt}} = 1\). We use Adam solver \cite{kingmaAdamMethodStochastic2017a} to initiate training of the networks from scratch, maintaining a consistent learning rate of 0.0001 for the initial 50 epochs. Subsequently, we linearly decrease the learning rate to zero over the following 50 epochs. Each epoch processed 10,000 images (we set the input dimension as \(256\times256 \, px\)) with a batch size of 1 (to perform instance normalization). The training was on a single Nvidia RTX A100 GPU; for dataset \hyperref[meth-micfl]{`GenLFI-Microfluidics'}, it took 29 hours. All tests were conducted on a workstation (CPU Intel i9-13900K, GPU Nvidia RTX3090). We utilized FID for quantification and comparative evaluation. Following fast reconstruction, we performed post-processing for large sample results using image stitching algorithms. As illustrated in Fig.\ref{fig-spheroids}\textbf{b}, we employed local object stitching \cite{weberAutomatischeZusammenfuehrungZertrennter2022} and global panorama stitching (see Supplementary Information 5) separately.

\subsection*{Cell spheroids imaging} \label{meth-sphimg}
Human pleural mesothelial cells MeT-5A NF2KO were cultured as 3D spheroids by seeding approximately 30,000 cells into each well of a 6-well ultra-low attachment culture plate (\#3471; Corning, Thermo Scientific); details see \cite{cunninghamYAPTAZActivation2023}. The dataset `GenLFI-Spheroids' construction and network training for this task follows a similar routine of  \hyperref[meth-micfl]{Microfluidics imaging}. We used a red laser (see \hyperref[meth-optset]{Optical setup}) as the light source. The bright-field microscope for recording is Leica DMIRE2 with a 10x objective lens, coupling with a polarization filter to enhance the light contrast for spheroid structure observation. The acquired raw data contains 132 frames of holograms and 26,581 frames of bright-field microscopic images. We constructed a dataset comprising 17,418 training and 100 test images (captured from an unseen well) for LFI holograms (after Bicubic interpolation super-resolution, \(808 \times 808 \, px\)), and 23,738 training and 100 test images for bright-field images (original size, \(1280\times1024 \, px\)). The time-lapse growth experiment was recorded for 2 hours at room temperature, as the spheroids needed to be cultured at \( {37}^\circ C\) in the incubator. 

\subsection*{Microfluidics imaging} \label{meth-micfl}
We fabricated a microfluidic chip with a flow-focusing geometry and conducted water-in-oil droplet generation experiments to observe these dynamic samples. The microfluidic channel width and height are \(200 \, \mu m\) and \(35 \, \mu m\), respectively. We used piezoelectric disc pumps (TTP Ventus HP series) to control the microfluidic flows and droplet production rates (ranging from 5 to 15 \(Hz\)). We conducted the imaging experiments under a bright-field microscope (Cerna modular microscopy with Thorlabs CS135MU camera) and our LFI platform, respectively, and acquired image data \(i\) and \(o\) for network training.

Using the GenLFI setup, we recorded 500 hologram frames (\(i\)) in total (.fit format, resolution \(6072\times4042 \, px\), acquired via the ASI Studio software) focusing on the ROI of the microfluidic chip with dynamic droplets (see Fig.\ref{fig-microfluidics}). Then, under the same experimental conditions,  we repeated the experiment to acquire real-world bright-field images (\(o\)) as the data guidance using a bright-field microscope. Due to the small FOV of the microscope, we scanned the same ROI and recorded a 5-minute video containing 7000 frames of bright-field images. 

We cropped the hologram images to patches with a size of \(166\times147 \, px\). Similarly, bright-field images were cropped to patches with \(1024\times1080 \, px\). The final dataset `GenLFI-Microfluidics' for this task comprises 121,886 training and 100 test cropped images from LFI hologram ROIs, and 57,193 training and 100 test cropped images from bright-field images. 


\subsection*{USAF 1951 test chart imaging} \label{meth-usaf}
We employed a USAF 1951 test chart (Thorlabs R3L3S1P) to evaluate the resolution limits of our LFI system. Given that the thickness of the test bench’s glass slide, \textit{i.e.,} \(1.5 \, mm\), affects the light diffraction, we placed the printed side close to the microscope's lens to minimize this effect when capturing images through standard bright-field microscopy. Then, we placed the sample in the same orientation to acquire holograms with GenLFI. For microscopy images, we selected a square ROI with the smallest pair of lines. For holograms, we cropped out the same ROI to a \( 220\times220 \, px\) image (see Fig.\ref{fig-resolution}\textbf{a}). We carefully aligned these two domain images on the same scale. To augment our dataset for this task, we introduced Gaussian noise to generate 10,000 variant images for each domain, of which 99 images with 1 original noise-free image as test images, and the other images as training images. 

\subsection*{Beads imaging} \label{meth-beads}
We used \(10 \, \mu m\) diameter beads solution (Invitrogen CML latex, \(4\%\) w/v, H2392891) diluted with water on a microscope slide for imaging. The water flow on the slide is approximately \(0.1 \, mm\) thick; therefore, the \(10 \, \mu m\) beads float or sink across the flow.  We recorded 23 frames of hologram images (\(6072\times4042 \, px\)) under the LFI platform, and 1430 frames (\(4096\times2160 \, px\)) images under the bright-field microscope with a 4x objective. For both recorded holograms and bright-field images, we applied Gaussian filtering to remove shadows, then cropped and applied data augmentation with vertical and horizontal flips to scale up the data size. The dataset, `GenLFI-Beads', consists of 96 test and 33,120 training images of hologram (\(506 \times505 \, px\)); 96 test and 32,576 training images of bright-field images (\(512\times540 \, px\)).

\bibliography{main}

\section*{Acknowledgement}\label{ackn}
R.L. was supported by the Global Joint PhD Studentship between KU Leuven and the University of Edinburgh. Y.Y. was supported by the European Research Council Starting Grant, Chancellor's Fellowship, and Bayes Innovation Fellowship. C.H.: The early stages of this work were supported by the JHMRF and Worldwide Cancer Research (19‐0238). K.P. is supported by an MRC Precision Medicine DTP Studentship. We thank Willemien Gosselé (KUL) and Mengguang Ye (UoE) for their assistance with the experiments.

\section*{Authors contributions}\label{contr}
R.L. contributed to all facets of this study, encompassing conceptualization, algorithm design, experimentation, analysis, interpretation, figure creation, and manuscript writing. Y.Y. and X.C. shared supervisory responsibilities for this project. P.B. supervised the optics experiments and offered valuable domain expertise in optics. Z.L. contributed to the panorama stitching algorithm. Additionally, P.B., M.W., K.P., G.F., Y.F., and C.H. contributed by providing samples and expertise for experimentation.

\section*{Competing interests}\label{comintr}
The authors declare no competing interests. 

\section*{Data and code availability}\label{dacoav}
Data and codes for the implementation are available at \url{https://github.com/RL-arch/LensGAN}.

\newpage
{\Large \textbf{Supplementary Information}}
\setcounter{figure}{0}
\renewcommand\thefigure{S\arabic{figure}} 

\section{Zebrafish imaging}\label{zebrafish}

\begin{figure}[h]%
\centering
\includegraphics[width=1.0\textwidth]{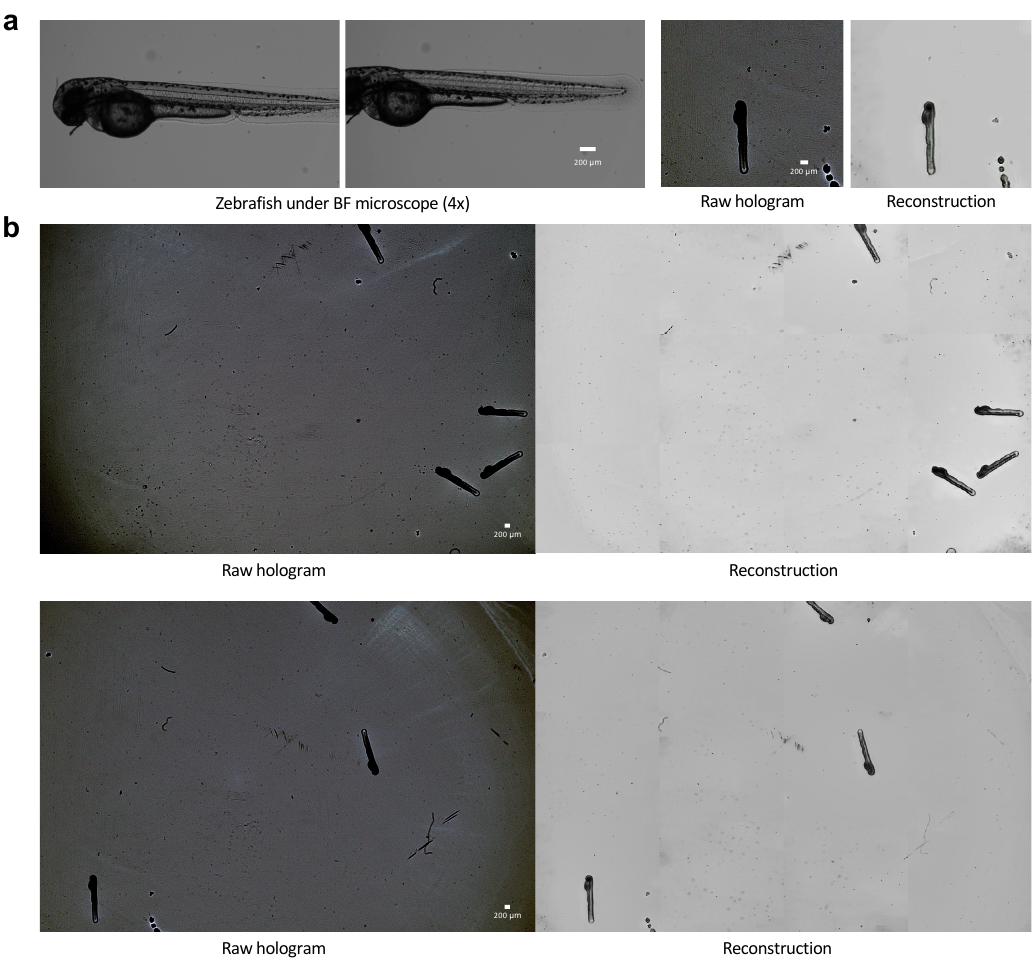}
\caption{ \textbf{a}, Details of imaging for zebrafish; \textbf{b}, Synthetized full FOV images. }\label{fig-fish}
\end{figure}

We applied GenLFI to in vivo samples of zebrafish. Zebrafish were cultured to 3 days old, anesthetized by MS222, and placed under petri dishes to be observed. Due to the structure of zebrafish, the measured hologram on our sensor (resolution \(5.94 \, \mu m/px\)) can provide very limited information inside the fish (see Fig.\ref{fig-fish}). Yet, our system can still reconstruct the zebrafish's basic structural information from raw holograms in a large FOV.

\section{Further discussion on complex optical field and 3D LFI}\label{optdiscussion}
\begin{figure}[H]
    \centering
    \includegraphics[width=\linewidth]{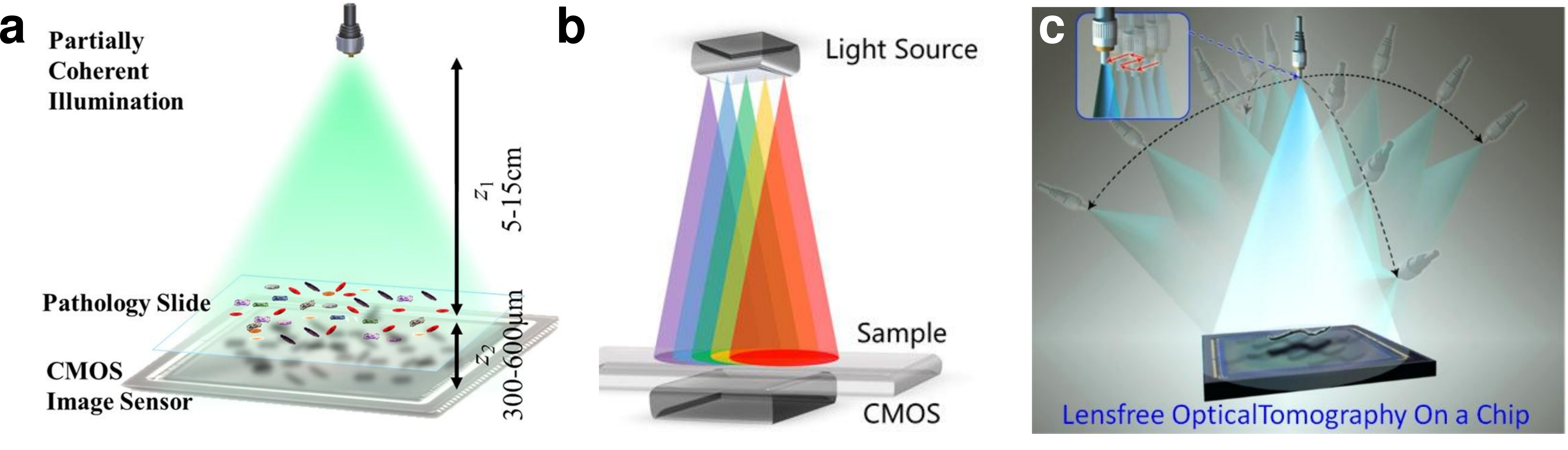}
    \caption[Multi‐measurement strategies in LFI]{Common multi‐measurement strategies for LFI.
    \textbf{a} Multi‐height phase recovery (MHPR) \cite{rivensonSparsitybasedMultiheightPhase2016}: captures holograms at several sample‐to‐sensor (\(z_2\)) distances. 
    \textbf{b} Multi‐wavelength retrieval \cite{wangMultiWavelengthPhaseRetrieval2022}: uses sequential illumination at five distinct wavelengths. 
    \textbf{c} Lens‐free 3D tomography \cite{isikmanLensfreeOpticalTomographic2011}: records holograms under varied illumination angles for volumetric reconstruction.}
    \label{fig:ch3_multimeas}
\end{figure}

To image 3D samples, modern digital in-line holography systems often employ measurement diversity strategies, as shown in Fig.\ref{fig:ch3_multimeas}. These include capturing multiple holograms by varying parameters such as sample-to-sensor distances \cite{rivensonSparsitybasedMultiheightPhase2016} and wavelengths \cite{wangMultiWavelengthPhaseRetrieval2022} in 2D LFI or illumination angles \cite{isikmanLensfreeOnChipMicroscopy2011} in 3D LFI. Furthermore, achieving high-resolution holograms frequently further relies on sub-pixel super-resolution via lateral (horizontal) shifting \cite{greenbaumImagingLensesAchievements2012, isikmanLensfreeOnChipMicroscopy2011}. Especially, lens-free 3D tomography (3D LFI, see Fig.\ref{fig:ch3_multimeas}c) is a 3D strategy for solving complex optical fields through multi-angle illumination.

\subsubsection*{Scalar diffraction theory}
One current solution for volumetric samples is 3D tomographic lens-free imaging (3D LFI), as illustrated in Fig.\ref{fig:ch3_multimeas}c. However, these methods, like FBP (filtered back-projection) \cite{isikmanLensfreeOpticalTomographic2011} or ODT (optical diffraction tomography) \cite{limComparativeStudyIterative2015}, also rely on the forward models. Here we discuss the optical field with large axial size samples and mathematically demonstrate the limitations of conventional model-based 3D lens-free holographic microscopy (LFHM) tomographic reconstruction approaches. To this end, we analyze the current 3D LFHM algorithm based on \textit{scalar diffraction theory} for 3D objects. 

Given a spatial vector  \(\mathbf(x\),  the unique solution to the light propagation problem is described as \cite{littlejohnPhysics221B2020}: 
\begin{equation}
    u(\mathbf{x}) = u^{inc}(\mathbf{x}) + u^{sct}(\mathbf{x})
    \label{eq:3Doptfield1}
\end{equation} 
where \(u^{inc}(\mathbf{x})\) represents the incident waves  and \(u^{sct}(\mathbf{x})\) represents the scattered waves, as shown in Fig.1b. The Lippmann-Schwinger equation (LSE) depicts the relationship between the 3D scattering potential of the sample and the resulting scalar field at region \( V(\mathbf{x})\): \begin{equation}
    u(\mathbf{x}) = u^{inc}(\mathbf{x}) + \int_V G(\mathbf{x}-\mathbf{x}')f(\mathbf{x}')u(\mathbf{x}')dr'
    \label{eq:3Doptfield2}
\end{equation}
where \(G(\mathbf{x})=\frac{1}{4\pi R}\exp(jk_{b}R)\) denotes Green's function for the unperturbed wave equation, \( k_{b} = k_{0}n_{b}\) is the wavenumber of the light in the background medium, \(f(\mathbf{x}) = k_0^2(n^2(\mathbf{x}-n_b^2)\) defines the scattering potential, and \(R\) denotes the length of the Euclidean vector \(\mathbf{x}\). With all variants discretized and denoted by vectors, the resulting scalar field \( \boldsymbol{u}\) can be represented by: 
\begin{equation}
    \boldsymbol{u} = \boldsymbol{u}^{inc} + \boldsymbol{G} \boldsymbol{\diag(f)} \boldsymbol{u}^{inc} = \mathcal{P}(\boldsymbol{f})
    \label{eq:3Doptfield3}
\end{equation}
where \(\mathcal{P}(\cdot)\) is a light propagator, \(\boldsymbol{\diag(f)} \in \mathbb{C}^{N \times N} \) denotes a diagonal matrix whose main diagonal entries consist of elements from the vector \( \boldsymbol{f}\). Here, \(\mathcal{P}(\boldsymbol{f})\) describes the optical filed crafted.

In LFI, only the intensity (amplitude) of \(\mathcal{P}(\boldsymbol{f})\), \textit{i.e.,} \(\left | \mathcal{P}(\boldsymbol{f})\right |^{2}\), from Eq.(\ref{eq:3Doptfield3}) is measured.  With multiple light sources, the final \textit{forward model} of the 3D LFHM can be presented by the intensity of \(\boldsymbol{H}\)  under \(k^{th}\) illumination: 

\begin{equation}
    \boldsymbol{H}_k = \left | \boldsymbol{u}^{inc} + \boldsymbol{G} \boldsymbol{\diag(f)} \boldsymbol{u}^{inc} \right |^{2} = \left | \mathcal{P}_{k} (\boldsymbol{f})\right |^{2}
   \label{eq:3Doptfield4}
\end{equation}

The reconstruction algorithm can be formulated as the optimization problem: 

\begin{equation}
    \boldsymbol{f}^{*} = \arg \min_{f \in \mathbb{C}^{N}} \lbrace \Sigma_{k=1}^{n} d(\left | \mathcal{P}_{k}(\boldsymbol{f})\right |^{2}, \boldsymbol{H}_{k})\rbrace,
    \label{eq:3Doptfield5}
\end{equation}

where \( \mathcal{P}_{k}(\cdot)\) denotes the light propagator, \(\boldsymbol{f}^{*}\) denotes the amplitude of the light, which is the measured resolvable hologram.

\subsubsection*{Boundary condition}
In the microfluidics experiment in Section \textit{Microfluidics screening}, the mediums consist of air, PDMS, water, oil, and glass. In this condition, different components of the electromagnetic field may make the vector diffraction theory of \(u(\mathbf{x})\) in Eq.(\ref{eq:3Doptfield1})  ineffective. Furthermore, the model-based algorithm's approximation of the hologram in the 2D space is affected by the \textit{thickness of the sample}. For example, due to the boundary condition of the continuity of the wave function and its derivative, the Born approximation for Eq.(\ref{eq:3Doptfield3}) is valid only if \(\boldsymbol{u}^{sct} \ll \boldsymbol{u}^{inc}\). This corresponds to \(l(n_{s} - n_{b}) \ll \lambda\) , where \(l\) is the sample thickness, \((n_{s} - n_{b})\)  describes the \textit{medium's refractive index variation}, and \(\lambda\) is the wavelength of the incident light. Consider a PDMS chip that has water inside, with a thickness of \(5000 \, \mu m\); there are \(l(n_{s} - n_{b})\) = \(5000 \times(1.40-1.33)= 350 \, \mu m\), while \(\lambda = 0.42 \, \mu m\).  It should become obvious that \textit{the model does not apply to the boundary condition of Born approximation} for large samples with intensive scattering. 

To accurately describe the physical process, some non-linear models have been proposed \cite{alma9993244227901488}.  These iterative methods involve either multiple iterations by variations or dividing the sample into a series of thin layers. However, these methods are impractical to apply in real experiments and are restricted to sample-specific tools. 

In summary, for \textit{dynamic 3D samples}, the real-world scattered fields are complex and challenging to model due to \textit{real-time variations in sample geometry and refractive index}. Current LFI tomography is therefore limited to static, and relatively thin samples.

\section{Full architecture of LensGAN}\label{fullnnarch}
\begin{figure}[h]%
\centering
\includegraphics[width=1.0\textwidth]{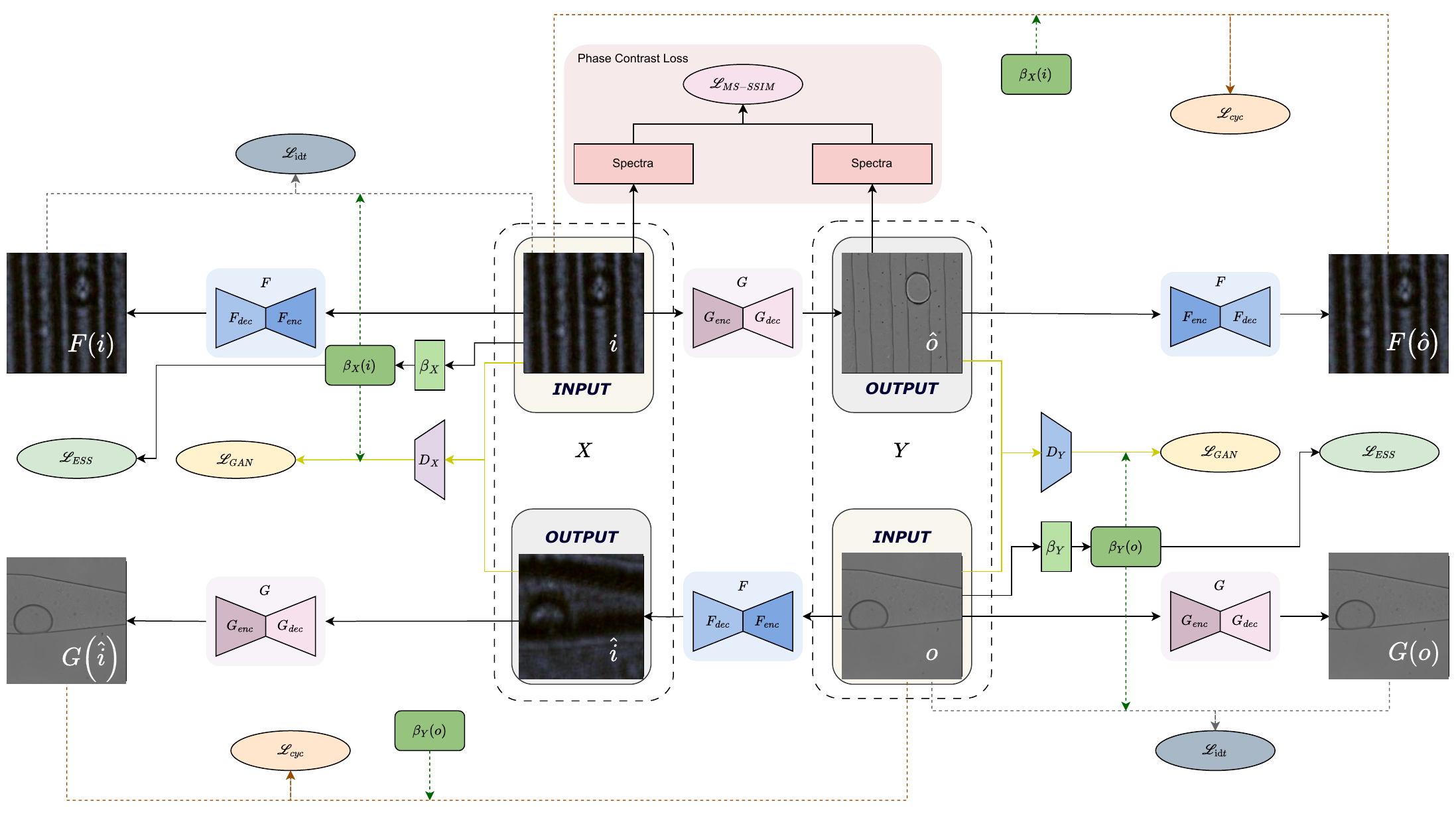}
\caption{The full architecture of LensGAN.}\label{fig-fullnn}
\end{figure}

\section{Discriminator design}\label{discriminator}

\begin{figure}[t]%
\centering
\includegraphics[width=0.6\textwidth]{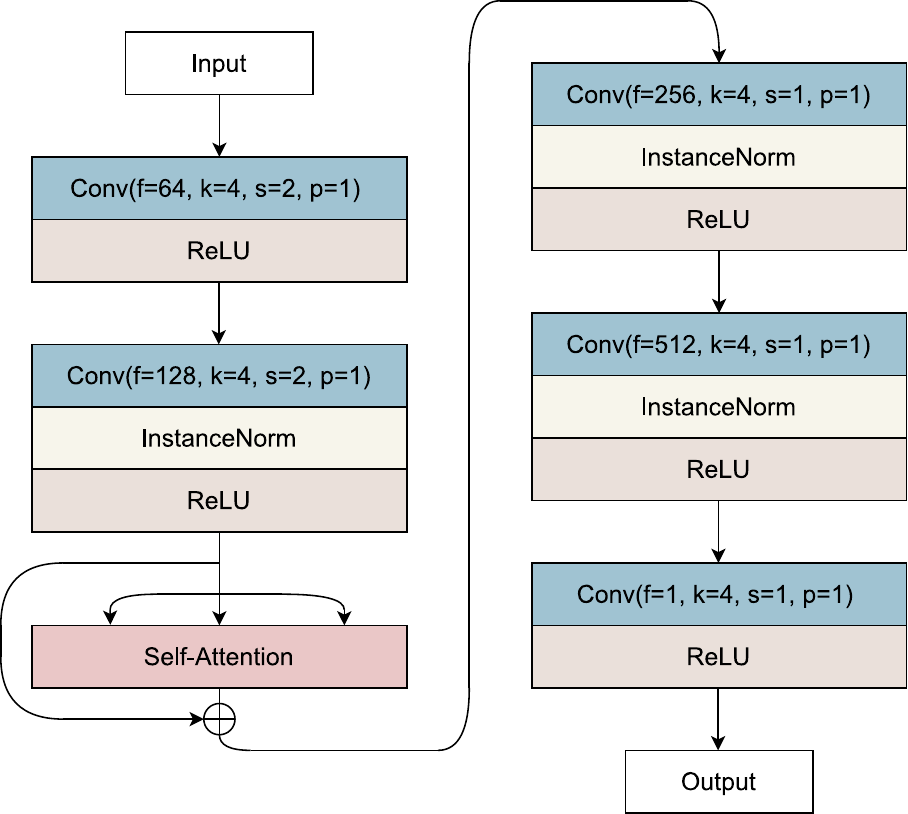}
\caption{ Architecture of the 5-convolutional-layer discriminator. f: filter, k: kernel size, s: stride, p: padding. The self-attention layer calculates the self-attention scores from 3 vectors: query, key, and value. Then it uses the scores to create an attended feature map that is added back to the input.  }\label{fig-dis}
\end{figure}

The discriminator network is based on \( 70 \times 70 \) PatchGANs \cite{isolaImagetoImageTranslationConditional2018}, which aims to classify whether \( 70 \times 70 \) overlapping image patches are real or fake. The input images under our hyperparameter setting are \( 256 \times 256 \), then downsampled by convolutional layers. The output is a one-channel prediction map to discriminate images. As shown in Fig.\ref{fig-dis}, the convolutional layers only process information in local neighborhoods. The sequential Markovian modeling of long-range dependencies in images may cause a limited receptive field. Therefore, a self-attention layer is added to solve this problem and make the training more robust. 

We implement this self-attention layer with the feature map used in SAGAN \cite{zhangSelfAttentionGenerativeAdversarial2019}.  The input image from the previous convolutional layer contains features \(x \in \mathbb{R}^{C \times N}\) that are transformed into two feature spaces $\mathnormal{f}$ and $\mathnormal{g}$. Here, $C$ represents the number of channels, and $N$ represents the number of feature locations. The attention function is calculated by the following equation where \( f(x) = W_{f}x \) , \( g(x) = W_{g} x \) : 
\begin{equation}
\beta_{j,i}=\frac{\exp(s_{ij})}{\sum_{i=1}^N\exp(s_{ij})}, \text{where } s_{ij} = f(x_{i})^Tg(x_{j})\label{eq:dis1}
\end{equation}
\( \beta_{j,i} \) indicates the range of attention where the model computes mapping of the \( i^{th}\) location of the \(j^{th} \) feature regions. The output of the attention layer is feature map \( a = (a_{1}, a_{2}, ...,a_{j},...,a_{N}) \in \mathbb{R}^{C \times N} \) , where
\begin{equation}
\begin{aligned}
     a_{j} = v(\sum_{i=1}^N \beta_{j,i} h(x_{i})) ,\\ h(x_{i}) = W_{h}x_{i}, \\ v(x_{i}) = W_{v}x_{i} \label{eq:dis2}
\end{aligned}
\end{equation}
\(W_{g} \in \mathbb{R}^{ \overline{C} \times C}\) , \(W_{f} \in \mathbb{R}^{\overline{C} \times C}\) , \(W_{h} \in \mathbb{R}^{\overline{C} \times C}\) , and \(W_{v} \in \mathbb{R}^{C \times\overline{C}}\) are three learned weight matrices that are implemented by \(  1 \times 1\) convolutions. Practically, for memory efficiency, the channel number of \( \overline{C}\) is reduced to be \( C/k\) , where \(k\) is set as 8. By multiplying \( a_{i}\) a learnable scale parameter \( \gamma \)  and adding back the input feature map \( x_{i}\),  the final scaled output is:
\begin{equation}
    y_{i} = \gamma o_{i} + x_{i} {.} \label{eq:dis3}
\end{equation}

\section{Post-processing: Global panorama image stitching} \label{imgstitch}
After processing the image blocks separately, the results will be stitched together to create the final panorama. This operation is divided into two stages. The first stage 
involves intensity correction and image assembly. The second stage aims to remove slowly varied artifacts.

For the first stage, considering two image blocks to be stitched, the color difference usually occurs between them, causing a noticeable seam. We define one image as the reference image $I^{r}$ and the other image as the target image $I^{t}$. Pixel points at the seam line are denoted by $p_i, i=1,2,...,n$. Then the color difference $\Delta C (p_i)$ between $I^{r}$ and $I^{t}$ at the seam line is defined as:
\begin{equation}
    \Delta C (p_i) = V^r(p_i) - V^t(p_i),~~i=1,2,...,n
\end{equation}
where $V^r(p_i)$ denotes the average pixel intensity at $p_i$ in the $I^{r}$ and $V^t(p_i)$ represents the average pixel intensity at $p_i$ in the $I_{t}$. $V^r(p_i)$ is calculated by averaging the pixel intensity at $p_i$ with its nearest three neighbors at the same row in the $I^{r}$. And the same approach is adopted to estimate $V^t(p_i)$. For each pixel $p^t$ in the $I^{t}$, the intensity change \(\Delta V\) is defined as \cite{fang2019fast}:
\begin{equation}
    \Delta V(p^t) = \frac{1}{\sum_{i=1}^n w_i} \sum_{i=1}^n w_i\Delta C (p_i),
\end{equation}
where the weight $w_i$ is expressed as:
\begin{equation}
   w_i = \mathrm{exp} \left(-\left(\frac{||V^t(p_i)-I^t(p^t)||^2}{\sigma_1^2}+\frac{||L(p_i)-L(p^t)||^2}{\sigma_2^2}\right)\right).
\end{equation}

Then, the corrected intensity at $p^t$ is formulated as:
\begin{equation}
 \overline{I^{t}(p^t)} = I^{t}(p^t)+\Delta V(p^t).
\end{equation}

There exist slowly varied artifacts in the generated panorama at the first stage. In the second stage, we treat this type of artifact as the low-frequency component of the image. Therefore, we can use the Gaussian filter to estimate these artifacts. Suppose the initial panorama is denoted by $P$, the artifact image is estimated by:
\begin{equation}
 A = P \ast \kappa,
\end{equation}
where $\ast$ represents the convolution, and $\kappa$ is the Gaussian kernel. Therefore, the final panorama $\overline{P}$ is expressed as:
\begin{equation}
\overline{P} = P - A.
\end{equation}

\begin{figure}[t]%
\centering
\includegraphics[width=1.0\textwidth]{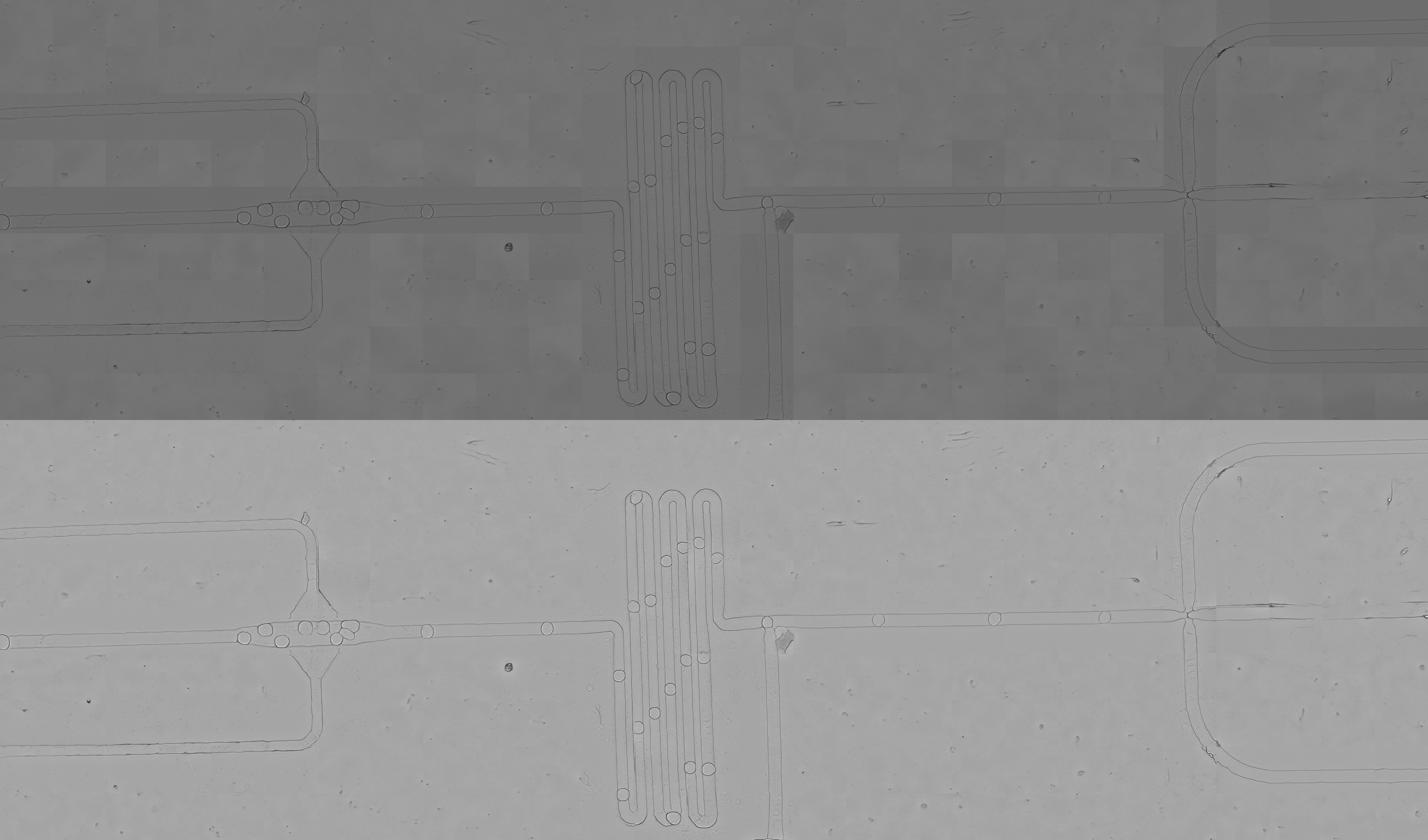}
\caption{Stitching result for microfluidics. Up: fast reconstruction. Down: after panorama stitching. }\label{fig-mfstitch}
\end{figure}

\begin{figure}[h]%
\centering
\includegraphics[width=1.0\textwidth]{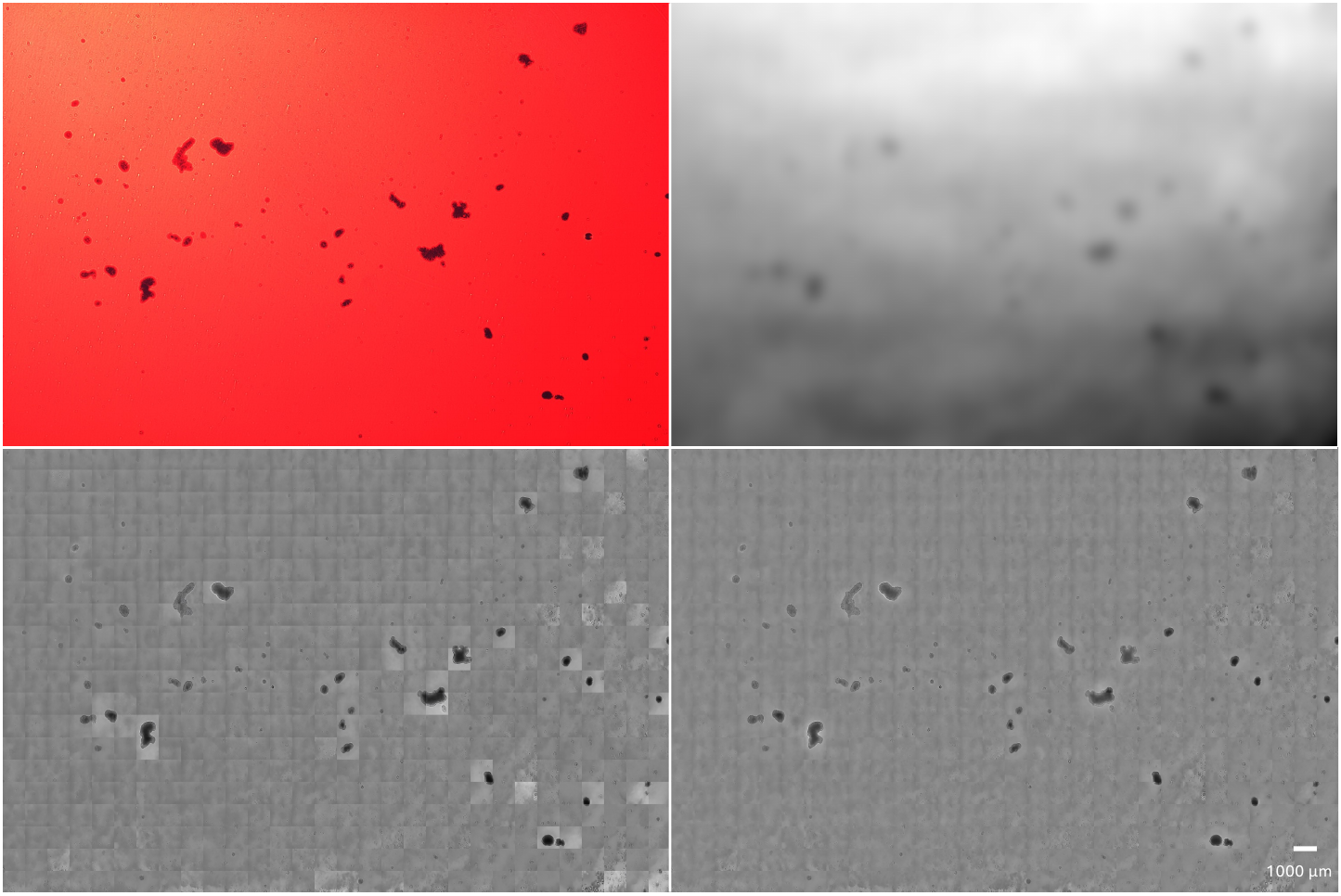}
\caption{Stitching result for spheroids. Top left: input raw hologram; Bottom left: fast reconstruction from LensGAN; Top right: detected shadow; Bottom right: after global panorama stitching and shadow removing.}\label{fig-spheroidstitch}
\end{figure}

\section{Microfluidic chip fabrication}\label{mffab}

We fabricated a microfluidic chip with a flow-focusing geometry for droplet production on PDMS bonded to a glass slide, and we conducted droplet generation experiments to observe these dynamic samples across a microfluidic system encompassing the droplet production area, a serpentine channel for droplet incubation and an expansion area for enabling droplet contact and merging. To fabricate the mold from which the microfluidic chip is made, photolithography is used. A \(35 \, \mu m\) thick layer of photoresist (SU-8 2015) is spin-coated, pre-baked, exposed, post-baked, and developed.

After the silicone wafer mold containing the design is created, a PDMS base and curing agent (Sylgard 184) are combined in a 10:1 ratio in a plastic cup and stirred for several minutes using a spatula. The plastic cup is placed in a vacuum desiccator to extract trapped air in the PDMS/curing agent mix. After approximately 20 minutes the mix was poured over the earlier created master, after which it was placed in the vacuum desiccator until no bubbles were visible anymore within the PDMS/curing agent mix. After this, the mix was placed in an oven at \(65^\circ\)C for 3 hours to cure the PDMS.

After the PDMS was cured, the PDMS was removed from the master, and separate devices were cut from this removed PDMS slab. Inlet and outlet holes were punched using a \(1.2 \, mm\) biopsy punch. Using scotch tape, dust was removed from both the glass slide and the PDMS devices, after which both were exposed to oxygen plasma in a plasma cleaner (Harrick PDC-002). The PDMS-coated glass slides and PDMS devices were then bonded by carefully connecting them.

Finally, the PDMS chip is silanized using a 5 w\% silane and HFE-7500 mixture.

\section{Spheroid genotypes} \label{orgagenotype}
The genotypes included are the wild type (WT) as well as two MeT5A clones with knockout (KO) of the \textit{NF2} gene – \textit{NF2} KO \#1 and \textit{NF2} KO \#2 (where \#1 and \#2 clones were generated using CRISPR constructs targeting different regions the gene). The \textit{NF2} gene has been found to have loss of function mutations in several cancers, including mesothelioma, and here these \textit{NF2} KO cell lines were generated as a cellular model system for mesothelioma. 


\section{FOV comparison}

\begin{figure}[h]%
\centering
\includegraphics[width=1.0\textwidth]{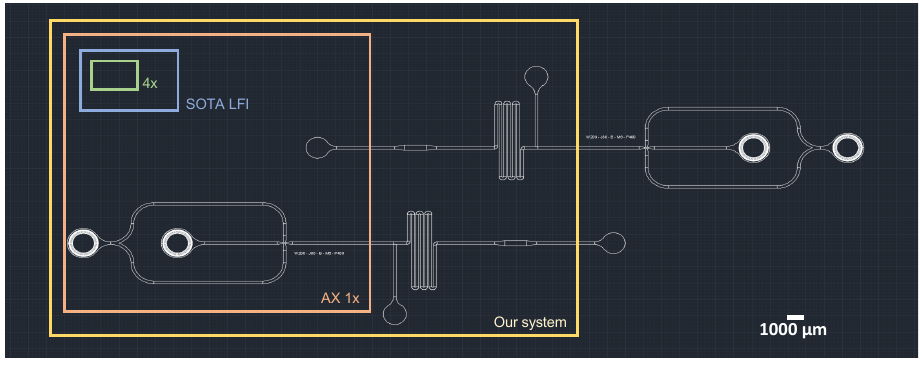}
\caption{FOV comparison, showing different imaging systems' FOV on our microfluidic chip design. 4x: 4x objective of the bright-field microscope; state-of-the-art LFI \cite{imecLensfreeImagingCompacta}: the existing LFI system's one-shot FOV based on holographic in-line reconstruction. AX 1x: Nikon AX \cite{AXAXNSPARC}, the world's largest commercial confocal microscope. The scanning 1x objective can cross a \(25\, mm\) diagonal FOV in multiple shots. With the low-resolution telescope setup, our FOV can reach \(> 500 \, mm^2\).}\label{fig-fovcompare}
\end{figure}

The FOV comparison of different imaging systems is shown in Fig.\ref{fig-fovcompare}.

\section{Shape Preservation Score (SPS) Evaluation}

To assess structural consistency between segmented objects in hologram and bright-field (BF) images from \textit{aligned} datasets, we introduce the \textbf{Shape Preservation Score (SPS)}. SPS quantifies the similarity of geometric shape descriptors between the two domains, measuring how well object-level morphological features are preserved during reconstruction.

\subsection*{Objective}

The goal of SPS is to evaluate the fidelity of shape reconstruction by comparing statistical distributions of key geometric descriptors computed from objects segmented in the hologram and BF domains.

\subsection*{Shape Descriptors}

We use three standard geometric descriptors \cite{lamThinningMethodologiesaComprehensive1992} to characterize object morphology:

\begin{enumerate}
\item \textbf{Aspect Ratio (AR)}: Measures the elongation of an object:
$     \text{AR} = \frac{L_{\text{major}}}{L_{\text{minor}} + \epsilon}
    $
where $L_{\text{major}}$ and $L_{\text{minor}}$ are the lengths of the major and minor axes of the ellipse fitted to the object. $\epsilon$ is a small positive constant (e.g., $10^{-6}$) added to prevent division by zero.

\item \textbf{Circularity (Circ)}: Reflects how close an object’s shape is to a perfect circle:
\[
\text{Circ} = \frac{4\pi A}{P^2 + \epsilon}
\]
where \( A \) is the object’s area and \( P \) its perimeter. Higher values indicate rounder shapes.

\item \textbf{Solidity (Sol)}: Quantifies the convexity of the object:
\[
\text{Sol} = \frac{A}{A_{\text{convex hull}}}
\]
where \( A_{\text{convex hull}} \) is the area of the convex hull enclosing the object. Solidity equals 1 for perfectly convex shapes.

\end{enumerate}

\subsection*{Metric Definition}

For each descriptor $d \in \{1, 2, 3\}$, we extract the empirical distributions $p_d(\text{holo})$ and $p_d(\text{bf})$ from all segmented objects in the hologram and BF images, respectively. We then compute the discrepancy between these distributions using the first-order Wasserstein distance ($W_1$, also known as Earth Mover’s Distance) \cite{arjovskyWassersteinGAN2017a}.

The overall SPS is defined as:

$$
\text{SPS} = \frac{1}{D} \sum_{d=1}^{D} W_1\left(p_d(\text{holo}),\ p_d(\text{bf})\right)
$$

where $D = 3$ is the number of descriptors.

\subsection*{Interpretation}

SPS provides a scalar measure of morphological consistency across imaging modalities. A lower SPS value indicates closer alignment of object shapes between the hologram and BF domains, implying better structural preservation. In contrast, a higher SPS suggests that shape features have been distorted during reconstruction, leading to domain mismatch.


\section{Fourier Ring Correlation (FRC) evaluation}

Fourier Ring Correlation (FRC) has become a popular and relatively unbiased method of estimating image quality/resolution recently, especially in localization-based superresolution microscopy \cite{kohoFourierRingCorrelation2019, riegerSingleImageFourier2024}. FRC  quantifies the correlation of spatial frequency content—specifically the magnitude of Fourier coefficients—between two images across concentric frequency bands, providing an objective measure of structural agreement at different resolutions.

\begin{itemize}
    \item Let $\mathcal{F}_1(u,v)$, $\mathcal{F}_2(u,v)$: Fourier transforms of the two images (shifted to center);
    \item Let $r(u,v) = \sqrt{(u - u_0)^2 + (v - v_0)^2}$ be the radial distance from center;
    \item For each radius $r_i$, define a ring $R_i$ (pixels where $r(u,v) = r_i$);
\end{itemize}

Then FRC at ring $i$ is:

$$
\text{FRC}(r_i) = \frac{\sum_{(u,v) \in R_i} |\mathcal{F}_1(u,v)||\mathcal{F}_2(u,v)|}{\sqrt{\sum_{(u,v) \in R_i} |\mathcal{F}_1(u,v)|^2 \cdot \sum_{(u,v) \in R_i} |\mathcal{F}_2(u,v)|^2} + \epsilon}
$$

The overall FRC score is:
$$
\text{FRC}_\text{mean} = \frac{1}{N} \sum_{i=1}^{N} \text{FRC}(r_i),
$$

where $N$ is the number of radial frequency bins (rings), $\epsilon$ denotes a small constant to prevent division by zero;

A higher FRC means better preservation of frequency content;
A score near 1.0 indicates a near-perfect correlation in Fourier magnitude structure.

\begin{figure}[t]
    \centering
    \includegraphics[width=0.5\linewidth]{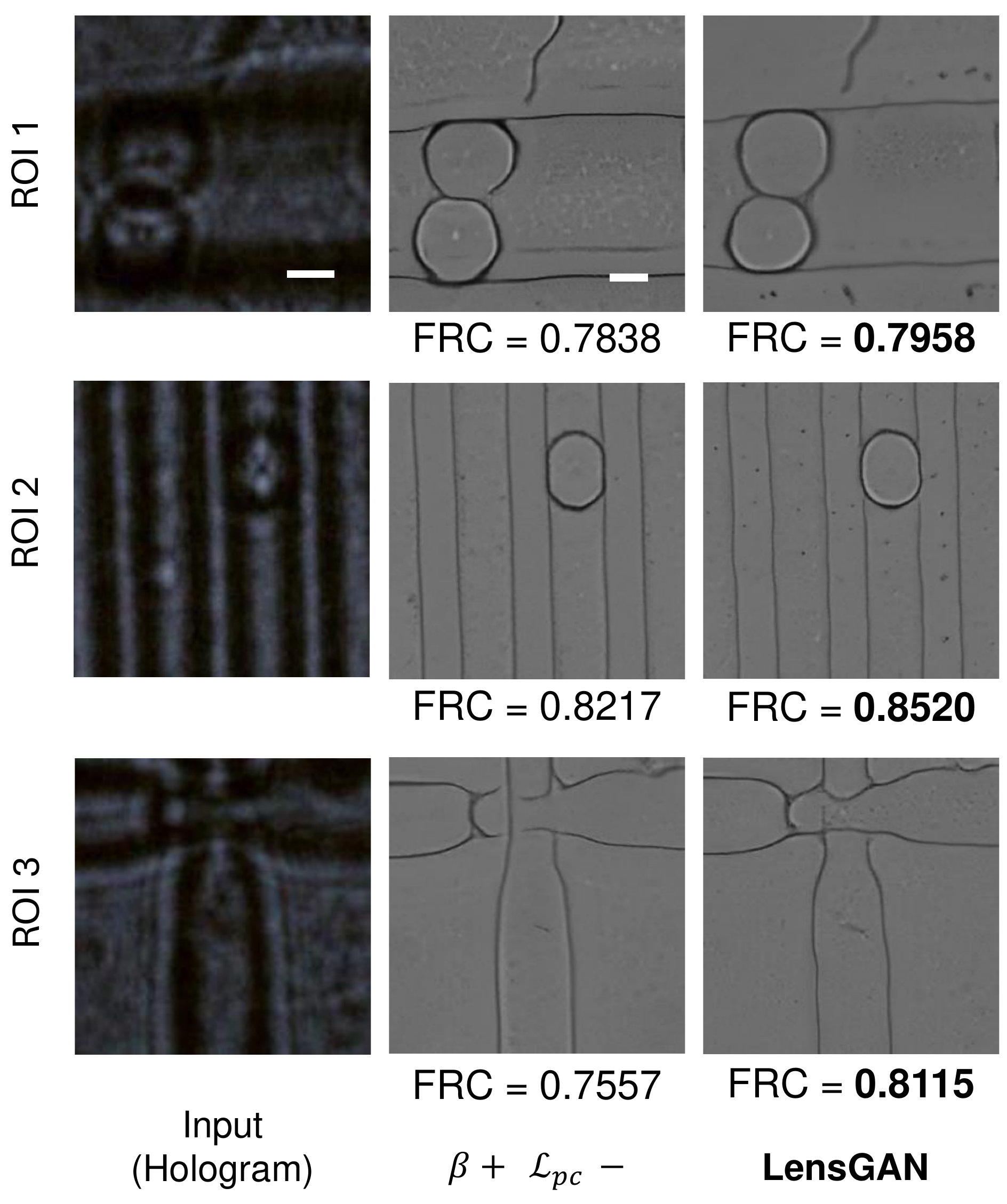}
    \caption{Supplementary to Fig 2, ablation study on phase-contrast loss evaluated by FRC. Scale bar: 100 \(\mu m\). }
    \label{fig:frc}
\end{figure}

We use FRC to evaluate the impact of the phase-contrast loss $\mathcal{L}_{\text{pc}}$. As shown in Fig.~\ref{fig:frc}, three ROIs from Fig. 2 in the main text were analyzed in an ablation study. With $\mathcal{L}_{\text{pc}}$, LensGAN consistently yields higher FRC scores across all ROIs, reflecting improved frequency-domain agreement and finer structural fidelity. This demonstrates that phase-contrast loss enhances the preservation of high-frequency spatial features, much like phase-contrast microscopy, which reveals subtle structural variations in transparent samples.

\section{Performance under noise conditions }

To evaluate the robustness of the model under varying noise conditions, we assessed LensGAN's performance under different noise levels. Specifically, with the same pretrained LensGAN model, we introduced additive Gaussian noise to the clean input images of the paired dataset \cite{triznaBrightfieldVsFluorescent2023}. 

\begin{figure}[h]
    \centering
    \includegraphics[width=\linewidth]{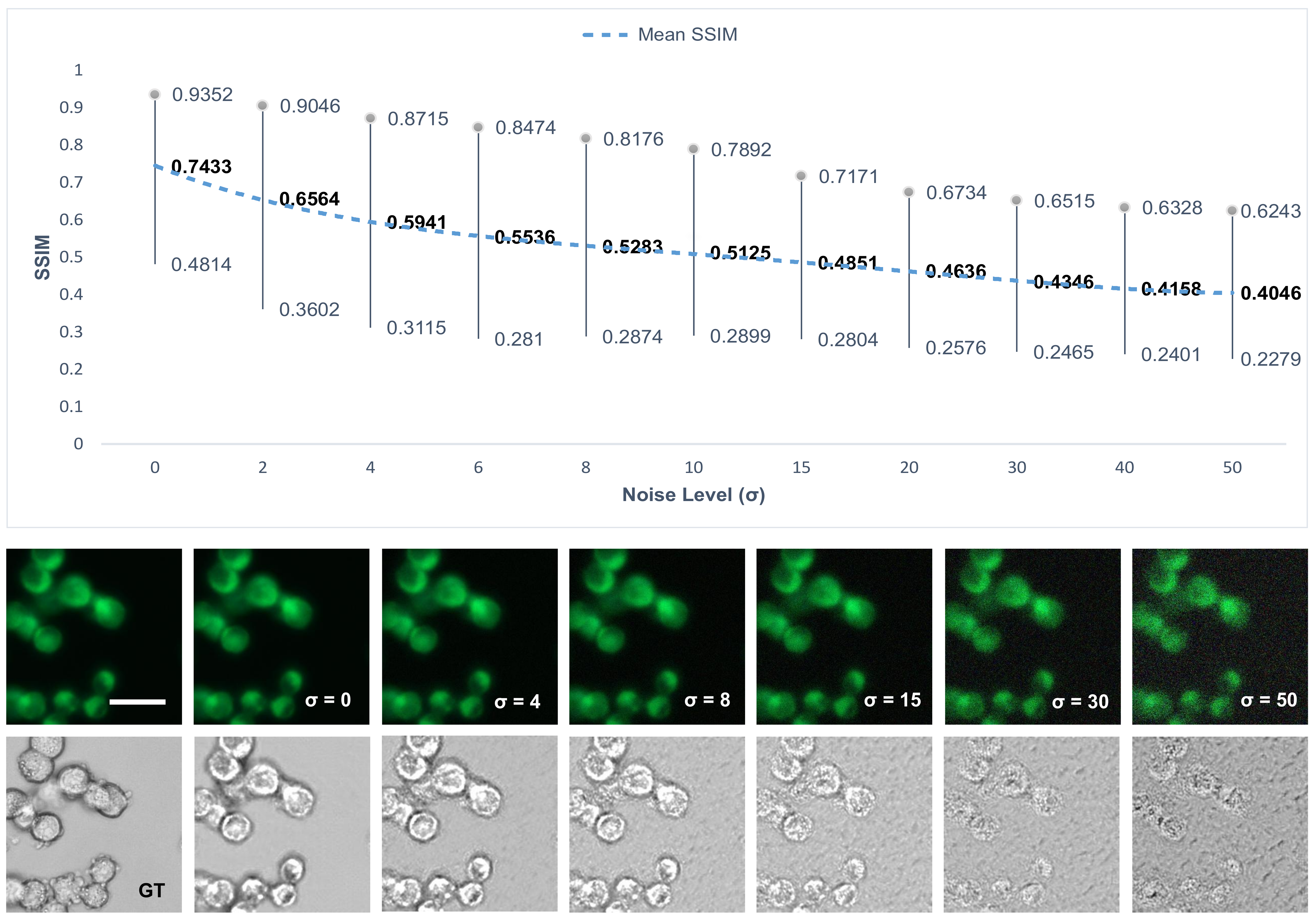}
    \caption{Performance under varying noise conditions are analyzed. The chart illustrates the lowest, highest, and mean average SSIM from 932 test sets.}
    \label{fig:noise-SSIM}
\end{figure}

Given an original image $I$, the noisy image $I_{\text{noisy}}$ is generated as:
$$
I_{\text{noisy}} = I + \mathcal{N}(0, \sigma^2)
$$

where $\mathcal{N}(0, \sigma^2)$ denotes Gaussian noise with zero mean and standard deviation $\sigma$. We tested noise levels at $\sigma \in \{2, 4, 6, 8, 10, 15, 20, 30, 40, 50\}$ to simulate a range of low to high noise conditions. The noise is applied pixel-wise, and values are clipped to the valid image intensity range (e.g., [0, 255] for 8-bit images). The results are illustrated in Fig.\ref{fig:noise-SSIM}. 

Our results demonstrate that image quality degrades consistently with increasing Gaussian noise levels. The SSIM mean decreased from 0.7433 at \(\sigma=0\) to 0.4046 at \(\sigma=50\), indicating a substantial loss of structural similarity under high noise of \(\sigma > 10\).  It is worth noting that the model was trained solely on clean, noise-free data; in practice, incorporating noise during training could substantially improve robustness to real-world imaging conditions.

\section{Multi-focus experiment}

\begin{figure}[h]
    \includegraphics[width=1\linewidth]{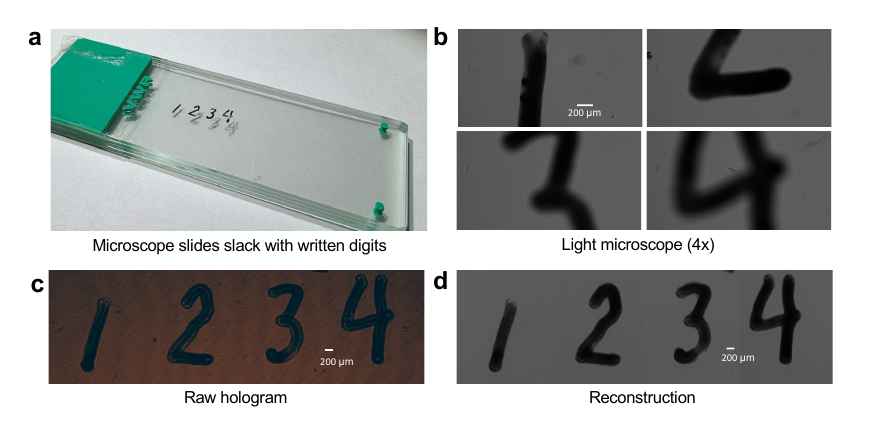}
    \caption[Multifocus]{Results of handwritten digits slides stack. Left shows the 4x objective focuses on “\textit{1}” and is gradually out of focus from “\textit{2}” to “\textit{4}”. Right is the reconstructed result; all numbers are focused.}
    \label{fig:multifocus}
\end{figure}

Unlike conventional light microscopes requiring manual focus adjustments, GenLFI excels at reconstructing in-focus images from holograms containing objects at different depths. This unique capability is attributed to its virtual depth of field, offering significant flexibility for multi-focal imaging. To showcase this advantage, we conducted experiments with two stacks of 4 glass slides (\(1 \, mm\) thick), each marked with a written digit (font-weight \(0.3 \, mm\)). In traditional bright-field setups, focusing is limited to a single depth plane, resulting in only one clear digit at a time (see Fig.\ref{fig:multifocus}). With proper data guidance, LensGAN can refocus and resolve all digits across different slides in a single reconstruction, achieving a superior depth of focus of 3 mm. This indicates our system's versatility and underscores its potential in scenarios where simultaneous multi-depth clarity is crucial. 

\end{document}